\documentclass[10pt]{article}

\input{./preamble.sty}

\usepackage[left=1in,top=1in,right=1in,bottom=1in,head=1.25in]{geometry}
\renewenvironment{abstract}
  {\small
   \begin{center}
   \bfseries \abstractname\vspace{-.5em}\vspace{0pt}
   \end{center}
   \list{}{%
     \setlength{\leftmargin}{10mm}%  <--- Adjust this value if you want
     \setlength{\rightmargin}{\leftmargin}%
   }%
   \item\relax}
  {\endlist}

\usepackage{dsfont,bm,color,appendix}
\usepackage{pdfpages}
\usepackage{algorithm}
\usepackage{algpseudocode}
\usepackage{bbm}
\usepackage[algo2e]{algorithm2e}
\usepackage{enumitem}
\usepackage[thinc]{esdiff}
\usepackage{afterpage}
\usepackage{amsmath,amssymb,bm}
\usepackage{booktabs}
\usepackage{multirow}
\usepackage{array}
\usepackage{xcolor}
\usepackage[symbol]{footmisc}
\usepackage[font=footnotesize,labelfont=bf]{caption}
\usepackage{lmodern}
\usepackage[square,numbers]{natbib}

\newcommand{\mR}{\mathcal{R}}

\newcommand{\be}{\begin{equation}}
\newcommand{\ee}{\end{equation}}
\renewcommand{\vec}[1]{\bm{#1}}

\newcolumntype{M}[1]{>{\centering\arraybackslash}m{#1}}

\newcommand{\closevdots}{\raisebox{5pt}{.}\raisebox{2.5pt}{.}\raisebox{0pt}{.}}

\newcommand{\bB}{{\boldsymbol B}}
\newcommand{\bD}{{\boldsymbol D}}
\newcommand{\bM}{{\boldsymbol M}}
\newcommand{\bV}{{\boldsymbol V}}
\newcommand{\bU}{{\boldsymbol U}}
\newcommand{\bW}{{\boldsymbol W}}

\newcommand{\bX}{{\boldsymbol X}}
\newcommand{\bZ}{{\boldsymbol Z}}
\newcommand{\bI}{{\boldsymbol I}}

\newcommand{\bSigma}{{\boldsymbol \Sigma }}
\newcommand{\bGamma}{{\boldsymbol \Gamma }}
\newcommand{\bOmega}{{\boldsymbol \Omega }}

\newcommand{\DifferenceMatrix}{\bD}
\newcommand{\Identity}{{\boldsymbol I}}
\newcommand{\balpha}{{\boldsymbol \alpha }}

\newcommand{\ag}[1]{{\bf{{\red{[{AG: #1}]}}}}}
\title{CoCA: Cooperative Component Analysis}
\author{Daisy Yi Ding$^{2*}$, Alden Green$^{1*}$, Min Woo Sun$^{2}$, and Robert Tibshirani$^{1,2}$  \\
$^{1}$Department of Statistics, Stanford University  \\ 
$^{2}$Department of Biomedical Data Science, Stanford University \\
*Equal contribution (alphabetical order)}

\begin{document}

\maketitle
\begin{abstract}
We propose \emph{Cooperative Component Analysis} (CoCA), a new method for unsupervised multi-view analysis: it identifies the component that simultaneously captures significant within-view variance and exhibits strong cross-view correlation.
The challenge of integrating multi-view data is particularly important in biology and medicine, where various types of ``-omic'' data, ranging from genomics to proteomics, are measured on the same set of samples.
The goal is to uncover important, shared signals that represent underlying biological mechanisms. 
%enabling a deeper understanding of biological systems and disease progression. %representative of the underlying mechanisms to understand the biology.
CoCA combines an approximation error loss to preserve information within data views and an ``agreement penalty'' to encourage alignment across data views.
By balancing the trade-off between these two key components in the objective, CoCA has the property of interpolating between the commonly-used principal component analysis (PCA) and canonical correlation analysis (CCA) as special cases at the two ends of the solution path. %spectrum. %encompasses both principal component analysis and canonical correlation analysis as special cases.
CoCA chooses the degree of agreement in a data-adaptive manner, using a validation set or cross-validation to estimate test error.
Furthermore, we propose a sparse variant of CoCA that incorporates the Lasso penalty to yield feature sparsity, facilitating the identification of key features driving the observed patterns.
We demonstrate the effectiveness of CoCA on simulated data and two real multiomics studies of COVID-19 and ductal carcinoma in situ of breast.
In both real data applications, CoCA successfully integrates multiomics data, extracting components that are not only consistently present across different data views but also more informative and predictive of disease progression. 
CoCA offers a powerful framework for discovering important shared signals in multi-view data, with the potential to uncover novel insights in an increasingly multi-view data world.

\end{abstract}

\section{Introduction}

With technological advances in biomedicine, it is common to collect multiple types of data, or ``data views'' on a common set of samples.
The multiple data views enable a more holistic characterization of the subjects under investigation.
For example, omics data, ranging from genomics and epigenomics to transcriptomics and proteomics, can now be routinely generated for a given set of biological specimens.
These omics data capture molecular variations from different dimensions, providing a comprehensive view of complex biological systems and offering the potential to uncover new insights that may be hidden in a single data modality.

Given the high dimensionality and complexity of these datasets, principal component analysis (PCA) is frequently used to reduce dimensionality and identify the most significant patterns within the data \cite{pearson1901liii, reich2008principal, alter2000singular, greenacre2022principal, novembre2008interpreting, allee2022pca, guellis2020pca, ghorbani2020pca, cunningham2014dimensionality}.
%(to do: add more?) \ag{Citations to justify?}.
However, in multi-view settings, the challenge extends beyond finding the principal components (PCs) that explain the most variance within data views. 
It can be equally important to ensure that the identified components exhibit a strong correlation across different data views.
This alignment of components across data views suggests that the uncovered patterns are not merely artifacts of a single data view but are consistent signals more representative of the underlying biology. 

On the other hand, canonical correlation analysis (CCA) is a commonly used method to identify patterns that are strongly correlated across different data views \citep{hotelling1992relations}.
It finds linear combinations of variables from two data views to maximize their correlation.
However, CCA has limitations, especially when dealing with high-dimensional data: it may be sensitive to noise and can pick up small noise directions \cite{witten2009penalized, gao2017sparse, witten2009extensions, janse2021cca}. 
%(to do: add more?) \ag{Citations to justify?}.
%may be misled by shared but unimportant factors that do not capture significant sources of variation in the data.
Additionally, CCA may overlook important signals as it does not take into account the variance explained by the identified components.

%NOTE: A picture here
%Algorithm producing the results
%A latent factor framework
%Another real data example, svd impute, compare imputation error, rho = 0 svd impute

\begin{figure}[h]
    \centering
    \includegraphics[width=\textwidth]{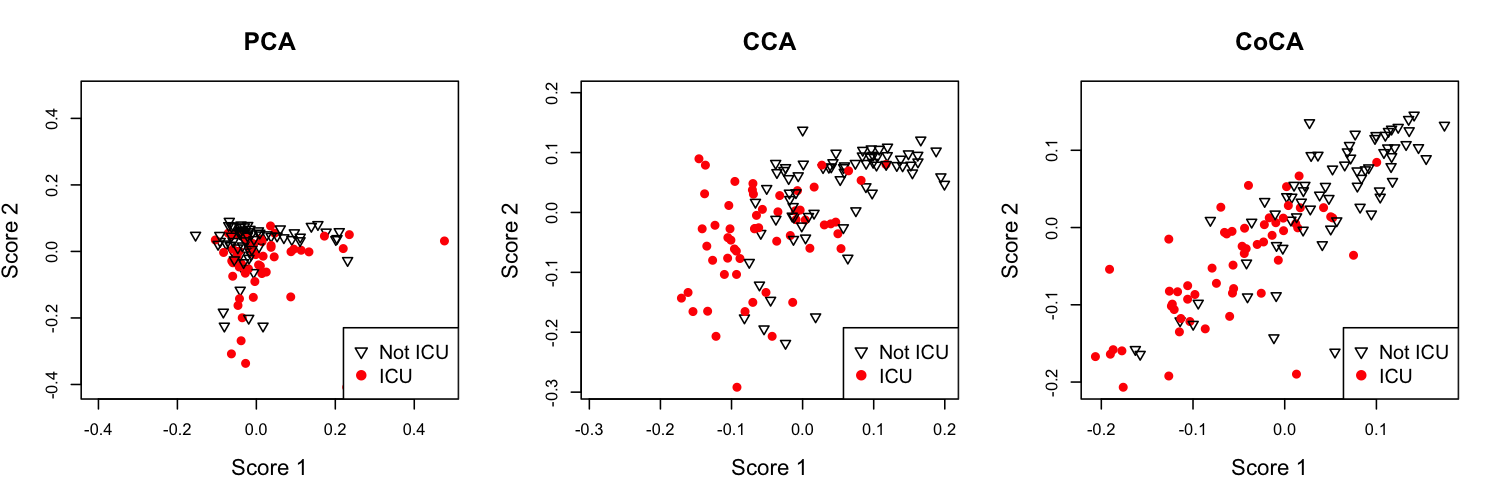}
    \caption{\emph{Comparison of multi-view scores derived from radiomics and clinical data views for a cohort of COVID-19 patients, using PCA, CCA, and CoCA, respectively.}
    %projections {\color{blue}(change here)} based on the components obtained from PCA, CCA, and CoCA for a cohort of COVID-19 patients, using radiomics and clinical data.} 
    Each point represents a patient, colored by ICU admission outcome. CoCA achieves clearer separation between ICU and non-ICU patients and better alignment between the two data views.}
    %Applying CoCA to a COVID-19 case study: The projected component scores for each patient based on the identified component are visualized as projections 1 and projections 2, corresponding to radiomics and clinical data views, respectively.}
    \label{fig:projections_intro}
\end{figure}

We propose a new method called \emph{Cooperative Component Analysis}, or short for CoCA, to identify the component that captures significant within-view variance and cross-view correlation in multi-view data.
To illustrate our method, we applied CoCA to integrate CT scan-derived radiomics features and clinical features measured on a cohort of 127 COVID-19 patients \citep{er2023multimodal}, which will be described in more detail in Section \ref{sec:real_data}. 
Figure \ref{fig:projections_intro} gives a visual illustration of the scores derived from radiomics and clinical data views, obtained from PCA, CCA, and CoCA, respectively. %onto the identified component
Each dot represents a patient and is colored by the patient's ICU admission outcome.
As compared to PCA and CCA, CoCA shows a clearer separation between patients who required ICU admission and those who did not, suggesting that it captures more informative patterns associated with COVID-19 disease progression. 
Moreover, CoCA also shows better alignment of the scores derived from the two data modalities, reflecting a shared underlying biological signal.
%which aims to take the best of both worlds by combining the variance-explaining power of PCA with the cross-view correlation-maximizing ability of CCA. 

The paper is organized as follows.
In Section \ref{sec:CoCA}, we introduce the formulation of CoCA and characterize its solution.
We show that it generalizes both PCA and CCA and further motivate the conceptual underpinnings of CoCA through an illustrative simulation study. 
%introduce a latent factor model to 
%We discuss its relation with PCA and CCA, 
%as well as other existing approaches.
%Furthermore, 
In Section \ref{sec:sparse_CoCA}, we present sparse CoCA that incorporates a Lasso penalty to improve the interpretability of the identified component, along with an efficient optimization algorithm for solving sparse CoCA.
We discuss its relation with other existing approaches and demonstrate its effectiveness in more simulation studies.
%in Section \ref{sec:simulation} the effectiveness of CoCA in simulation studies. 
In Section \ref{sec:real_data}, we illustrate the practical utility of CoCA through its application to two real-world multiomic datasets: (1) integration of radiomics and laboratory measurements of COVID-19 patients, and (2) integration of epithelial and stromal gene expression data of breast ductal
carcinoma in situ patients.
In both applications, we show that CoCA uncovers components that capture significant signals within data views while exhibiting strong correlations across views.
%We integrate radiomics and clinical features to uncover the component that captures important signals within data views and shows strong correlations across views.
%the shared components between these two data views. 
%We demonstrate that CoCA not only achieves better performance in identifying the relevant component but also leads to improved downstream prediction of patients' risks of severe disease progression requiring ICU admission.
%In Section \ref{sec:extension}, we discuss how CoCA can be extended to multiple components.
%and to settings where explicit grouping information is unavailable or suboptimal.
The paper ends with a discussion in Section \ref{sec:discussion} and an Appendix.
%{\color{blue} Update this paragraph after all the changes}

\section{Cooperative component analysis (CoCA)}
\label{sec:CoCA}

\subsection{CoCA}
We begin with a simple form of cooperative component analysis (CoCA) without sparsity.
Let ${\boldsymbol X}_1 \in \Reals^{n \times p_1}, {\boldsymbol X}_2 \in \Reals^{n \times p_2}$ --- representing two data views --- and ${\boldsymbol X} \in \Reals^{n \times p}$ be the concatenation of ${\boldsymbol X}_1$ and ${\boldsymbol X}_2$.
Fixing the hyperparameter $\rho \geq 0$, we propose to minimize the following objective:

\begin{equation}
    \min_{u,v,d} \frac{1}{2}\|\bX - d u v^{\top}\|_{F}^2 + \frac{\rho}{2}  \|d\bX_1 v_1 - d\bX_2 v_2\|_2^2 , \quad \st \|v\|_2^2 = 1, \|u\|_2^2 = 1,
    \label{eqn:cooperative-svd}
\end{equation}

where $u \in \Reals^{n}$ and $v \in \Reals^{p}$ are vectors, and $d$ is a scalar.
$v$ is partitioned as $v = (v_1, v_2)$, where $v_1  \in \Reals^{p_1}$ corresponds to $\bX_1$ and $v_2 \in \Reals^{p_2}$ to $\bX_2$. %\ag{Say what dimension all of these are.}
The rows of the data matrices $\bX_1$ and $\bX_1$ represent the common set of observations, and the columns correspond to features.
The objective of CoCA is comprised of two key components:
\vspace{3mm}
\begin{enumerate}
    \item \textbf{Approximation error}: $\|\bX - d u v^{\top}\|_{F}^2$ measures the Frobenius norm of the difference between the original data ${\bX}$ and its low-rank approximation $d u v^{\top}$. %\ag{Witten et al define reconstruction error to be $\|\bX - \bX u v^{\top}\|_{F}^2$. I might call this low-rank approximation error, or just approximation error.} 
    Specifically, given a data matrix $\bX \in \Reals^{n \times p}$ with singular value decomposition (SVD) $\bX = \sum_{k = 1}^{r} d_k u_k v_k^{\top}$, by the Eckart-Young Theorem \cite{eckart1936approximation}, the leading singular value $d_1$, and left and right singular vectors $u_1,v_1$, are the solution to the optimization problem 
    %\ag{I would say ``by the Eckart-Young Theorem...'' or somethign like that}
    \begin{equation}
        \min_{u,v,d} \|\bX - d u v^{\top}\|_{F}^2, \quad \st \|v\|_2^2 = 1, \|u\|_2^2 = 1.
    \end{equation}
    The constraints $\|v\|_2^2 = 1$ and $\|u\|_2^2 = 1$ ensure the normalization of the singular vectors to guarantee a unique solution.  Note that PCA can be performed via SVD of the data matrix: the principal components are the right singular vectors of the data matrix obtained from SVD. %\ag{Check: I think they ensure a unique solution rather than avoiding trivial solutions.}
    \vspace{3mm}
    \item \textbf{Agreement Penalty}: $\|d\bX_1 v_1 - d\bX_2 v_2\|_2^2$ introduces a penalty to encourage alignment across different data views. Intuitively, this term aligns two sets of scores for the samples, obtained by computing $\bX_{1}v_{1}$ and $\bX_{2}v_{2}$. %to encourage the projections from different data views to align, therefore promoting agreement across data views
    %projecting $\bX_1$ onto $v_1$ and $\bX_2$ onto $v_2$. \ag{Careful, $\bX_1v_1$ and $\bX_2v_2$ are not projections. The right word for them is scores, as you use in the following.} 
    These scores could represent, for example, disease severity measures, derived from radiomics and clinical data views as in the previous motivating example. 
    We refer to these as multi-view scores. 
    By minimizing the difference between these two sets of scores, the agreement penalty encourages similarity between the disease severity measures obtained from radiomics data and those from clinical data, thereby promoting agreement across data views.
    %By minimizing the difference between these two sets of scores, the agreement penalty leverages the shared underlying relationships across views, ensuring that the identified components capture consistent patterns in the multi-view data.\ag{Last sentence may be a little too imprecise. An idea: why don't you carry on with the nice motivating example of the previous sentence? E.g., if we have two measures of disease severity, one from radiomics data and one from clinical data, the agreement penalty encourages them to be similar.}
    $\rho \geq 0$ controls the relative importance of the agreement penalty.
    %onto the corresponding components to align, thus encouraging agreement across data views. 
    %This penalty enforces similarity between the projections of ${\bX}_1$ and ${\bX}_2$ onto their respective singular vectors $v_1$ and $v_2$, thus encouraging agreement across the views.
\end{enumerate}
\vspace{3mm}

%It works by combining two key components in the objective: a reconstruction error loss and an ``agreement'' penalty. 
%The reconstruction error loss quantifies the discrepancy between the original data and its low-rank approximation based on the identified component so that it captures as much within-view information as possible.
%reconstructed version from a lower-dimensional space.
%mirroring PCA's objective to retain as much variance as possible in the reduced-dimensional representation.
%Simultaneously, the agreement penalty encourages the projections from different data views onto the corresponding components to align, thus encouraging cross-view agreement.

CoCA formulation reflects a balance between capturing important patterns within data views and encouraging alignment across data views to reveal common underlying signals. 
The hyperparameter $\rho$ controls the trade-off between the approximation error and the agreement penalty. 
When $\rho = 0$, Problem~\ref{eqn:cooperative-svd} reduces to the standard SVD, which focuses solely on minimizing the approximation error. 
%\ag{Same comment re: reconstruction error.}
As $\rho$ increases, more emphasis is placed on encouraging alignment across data views.
The optimal value of $\rho$ can be estimated using a validation set or through cross-validation, which we describe in more detail in Appendix Section \ref{sec:appendix_cv}. %\ag{Do we explain how to do cross-validation somewhere? If so, make a reference to it. If not, we should.}

By varying the weight of the agreement penalty, CoCA has the property of encompassing PCA and CCA as special cases at the two extremes of the solution path:
\begin{itemize}
    \item When \textbf{$\rho = 0$}, the solution $\hat{v}$ is proportional to the first principal component of the combined data view $\bX$;
    \item As $\rho$ approaches infinity, the appropriately scaled $\hat{v}$ converges to the leading canonical direction between the two views $\bX_{1}$ and $\bX_{2}$.
\end{itemize}
We discuss this relationship in more detail in the next section.
%Add rho, PCA and CCA
%In Section \ref{sec:pca_cca}, we discuss the relationship of CoCA to PCA and CCA when varying the weight of the agreement penalty.
%or other model selection techniques based on the specific application and desired balance between within-view variance explanation and cross-view correlation.
	
\subsection{Relation to PCA and CCA}
\label{sec:pca_cca}

We now show that CoCA generalizes both PCA and CCA: when \textbf{$\rho = 0$}, $\hat{v}$ is proportional to the first principal component of $\bX$, and as \textbf{$\rho \to \infty$}, the appropriately scaled $\hat{v}$ converges to the leading canonical direction.
We will find it convenient to transform $dv \mapsto v$ in Problem~\eqref{eqn:cooperative-svd}, in which case the problem can be reformulated as\footnote{Note that if $(\hat{u},\hat{v})$ is the solution to~\eqref{eqn:cooperative-svd-2}, then $(\hat{u},\hat{v}/\|\hat{v}\|_2,\|\hat{v}\|_2)$ is the solution to~\eqref{eqn:cooperative-svd}.}
\begin{equation}
    \label{eqn:cooperative-svd-2}
    \min_{u,v} \frac{1}{2}\|\bX - u v^{\top}\|_{F}^2 + \frac{\rho}{2} \|\bX_1 v_1 - \bX_2 v_2\|_2^2 \quad \st \|u\|_2^2 = 1.
\end{equation}

Let $\hat{u},\hat{v}$ be the solution to~\eqref{eqn:cooperative-svd-2}. 
Standard Lagrange calculus shows that $\hat{u}$ is the leading eigenvector of the matrix $\bX (\Identity + \rho \bD \bX^{\top} \bX \bD)^{-1} \bX^{\top}$, while $\hat{v}$ is the leading eigenvector of the matrix $(\Identity + \rho \bD \bX^{\top} \bX \bD)^{-1} \bX^{\top} \bX$, where $\bD = \text{diag}(\Identity_{p_1}, -\Identity_{p_2})$.
Let $\lambda_1$ be the leading eigenvalue of $(\Identity + \rho \bD \bX^{\top} \bX \bD)^{-1} \bX^{\top} \bX$.
\begin{theorem}
\label{thm:coca-solution-path}
The solution~$(\hat{u},\hat{v})$ to~\eqref{eqn:cooperative-svd-2} satisfies
\begin{equation} 
\label{eqn:cooperative-svd-u} 
\bX (\Identity + \rho \bD \bX^{\top} \bX \bD)^{-1} \bX^{\top} \hat{u} = \lambda_1 \hat{u}, \;\; \|\hat{u}\|_2^2 = 1. 
\end{equation} 
and
\begin{equation} 
\label{eqn:cooperative-svd-v} 
(\Identity + \rho \bD \bX^{\top} \bX \bD)^{-1} \bX^{\top} \bX \hat{v} = \lambda_1 \hat{v}, \;\; \|\bX \hat{v}\|_2 = \lambda_1. 
\end{equation} 
Moreover, when $\rho = 0$, $\hat{v}$ is proportional to the first principal component of $\bX$. As $\rho \to \infty$, if $\rank(\bX) = p < n$ and $\bX_1^{\top} \bX_2 \neq 0$, then $(\hat{v}_1/\|\bX_1\hat{v}_1\|_2,\hat{v}_2/\|\bX_2\hat{v}_2\|_2)$ converges to the first canonical direction between $\bX_1$ and $\bX_2$.
\end{theorem}
The proof of Theorem~\ref{thm:coca-solution-path} is given in Appendix~\ref{sec:appendix_relation}. An interesting consequence of Theorem~\ref{thm:coca-solution-path} is that CoCA can be viewed from two perspectives:
\begin{enumerate}
    \item The perspective we have been taking thus far is that CoCA is a kind of \emph{penalized PCA}. The CoCA objective takes the traditional objective of SVD -- minimizing rank-1 approximation error -- and adds a penalty that encourages the solution to be highly correlated between views. If the first population PC is highly correlated between views, the agreement penalty exploits this structure so that the CoCA solution can be more accurate, and achieve lower reconstruction error on test data, than the usual first sample PC. 
    \item A new perspective, motivated by Theorem~\ref{thm:coca-solution-path}, is that CoCA is a kind of \emph{penalized CCA}. Seen in this way, CoCA takes an objective -- minimizing squared differences between the single-view scores $\bX_1 v_1$ and $\bX_2 v_2$ -- that results in the first canonical directions, and adds a penalty that steers the solution towards higher variance-explained solutions. This penalty prevents overfitting spurious correlations in lower variance-explained directions. If $\bX_1,\bX_2$ have low effective rank, and the first population canonical directions are indeed in the higher variance-explained subspaces, then the CoCA solution can be more accurate than the usual first sample canonical directions.
\end{enumerate}
These two perspectives on CoCA are visualized in Figure~\ref{fig:CoCA}. The upshot is that CoCA can be viewed as an alternative to either PCA (for estimating the first population PC) or CCA (for estimating the first canonical direction), and we will see examples where CoCA is more accurate than either PCA or CCA shortly.

% An interesting consequence of Theorem 2 is that CoCA can be viewed from two perspectives:
%\begin{enumerate}
%    \item On the one hand, it can be seen as finding the best rank-1 approximation of the data while leveraging side information, i.e. shared underlying relationships between views, to improve approximation error (panel (1) in Figure \ref{fig:CoCA}). The assumption is that the best rank-1 approximation is the information that is redundantly expressed across views, and the agreement penalty exploits this structure to improve the approximation.
%    \item Alternatively, CoCA can be interpreted as identifying the direction that maximizes correlation across views while incorporating a constraint that steers the solution towards high-variance directions and reduces overfitting  (panel (2) in Figure \ref{fig:CoCA}). This additional constraint helps prevent the method from being misled by small noise components, a potential drawback of traditional CCA.
% \end{enumerate}
% \fi

\subsection{Illustrative simulation study}
To show the potential benefits of CoCA, as compared to PCA and CCA, we conduct an illustrative simulation study with data drawn from a latent factor model. In this model, we suppose the data $\bX = (x_1 \dots x_n)^{\top}$ consist of independent vectors $x_i \in \R^p$, with each $x_i = (x_{i,1}, x_{i,2})$ consisting of two views $x_{i,1} \in \R^{p_1}, x_{i,2} \in \R^{p_2}$ generated according to
\begin{equation}
\begin{aligned}
\label{eqn:multi-view-latent-space-model}
x_{1} & = \beta_1 z + \bW_1 z_1 + \bB_1 s + \epsilon_{1}, \quad z \sim N(0,1), z_1 \sim N_{k_1}(0,\bI_{k_1}), s \sim N_{l}(0, \bI_{l}), \epsilon_{1} \sim N_{p_1}(0,\bOmega_1)  \\
x_{2} & = \beta_2 z + \bW_2 z_2 + \bB_2 s +  \epsilon_{2}, \quad z_2 \sim N_{k_2}(0,\bI_{k_2}), \epsilon_{2} \sim N_{p_2}(0,\bOmega_2)
\end{aligned}
\end{equation}
with all random variables independent. Latent variable models similar to~\eqref{eqn:multi-view-latent-space-model} have been used to study PCA~\cite{johnstone2001distribution,tipping1999probabilistic,paul2007asymptotics,johnstone2009consistency} and CCA~\cite{bach2005probabilistic,chen2013sparse}, and so~\eqref{eqn:multi-view-latent-space-model} is a natural test bed for understanding CoCA. The covariance of a random vector $x \in \R^p$ distributed according to~\eqref{eqn:multi-view-latent-space-model} is
$$
\mathrm{Cov}(x) := \bSigma = \beta \beta^{\top} 
+ \begin{bmatrix} 
\bW_1 \bW_1^{\top} & {\boldsymbol 0} \\
{\boldsymbol 0} & \bW_2 \bW_2^{\top}
\end{bmatrix} 
+ \bB \bB^{\top} + 
\begin{bmatrix}
\bOmega_1 & {\boldsymbol 0} \\
{\boldsymbol 0} & \bOmega_2
\end{bmatrix},
$$
where $\beta = (\beta_1,\beta_2)$ and $\bB = (\bB_1,\bB_2)$.

Consider~\eqref{eqn:multi-view-latent-space-model}, with $p_1 = p_2 = 4$, and parameters taken as follows:
$$
\beta_1 = \beta_2 = \begin{bmatrix} 1 \\ 0 \\ 0 \\ 0 \end{bmatrix}, \; \bW_1 = \bW_2 = (\|\beta\|_2 - 0.1)\begin{bmatrix}0 \\ 1 \\ -1 \\ 0\end{bmatrix},\; \bB_1 = \bB_2 = (\|\beta\|_2 - 1) \begin{bmatrix}0 \\ 0 \\ 0 \\ 1\end{bmatrix},\; \bOmega_1 = \bOmega_2 = \diag(1, 1, 1, 0.09). 
$$
For these choices, we would expect an intermediate value of $\rho$ to achieve the best performance:
\begin{itemize}
    \item The first population PC and population canonical directions are both equal to $\beta = (\beta_1,\beta_2)$. Thus $\beta$ corresponds to the direction that both explains the highest variance, and has the most cross-correlation between views. It is in fact also the solution to CoCA at the level of the population -- i.e. the optimization problem~\eqref{eqn:cooperative-svd} with $\bX = \bSigma^{1/2}$ -- for any value of $\rho$, as we demonstrate numerically in this simulation study. 
    \item Both $(\bW_1,0)$ and $(0,\bW_2)$ -- i.e. $\bW_1$ and $\bW_2$, padded with zeros to make them vectors in $\Reals^p$ -- explain a similar amount of variance to $\beta$, making them good candidates for PCA; but they are each supported in only one of the two views, and so have cross-correlation equal to zero.
    \item On the other hand, $\bB$ has similar cross-correlation to $\beta$, making it a good candidate for CCA; but it explains much less of the overall variance.
\end{itemize}

\begin{figure}[t]
    \centering
    \includegraphics[width=\textwidth]{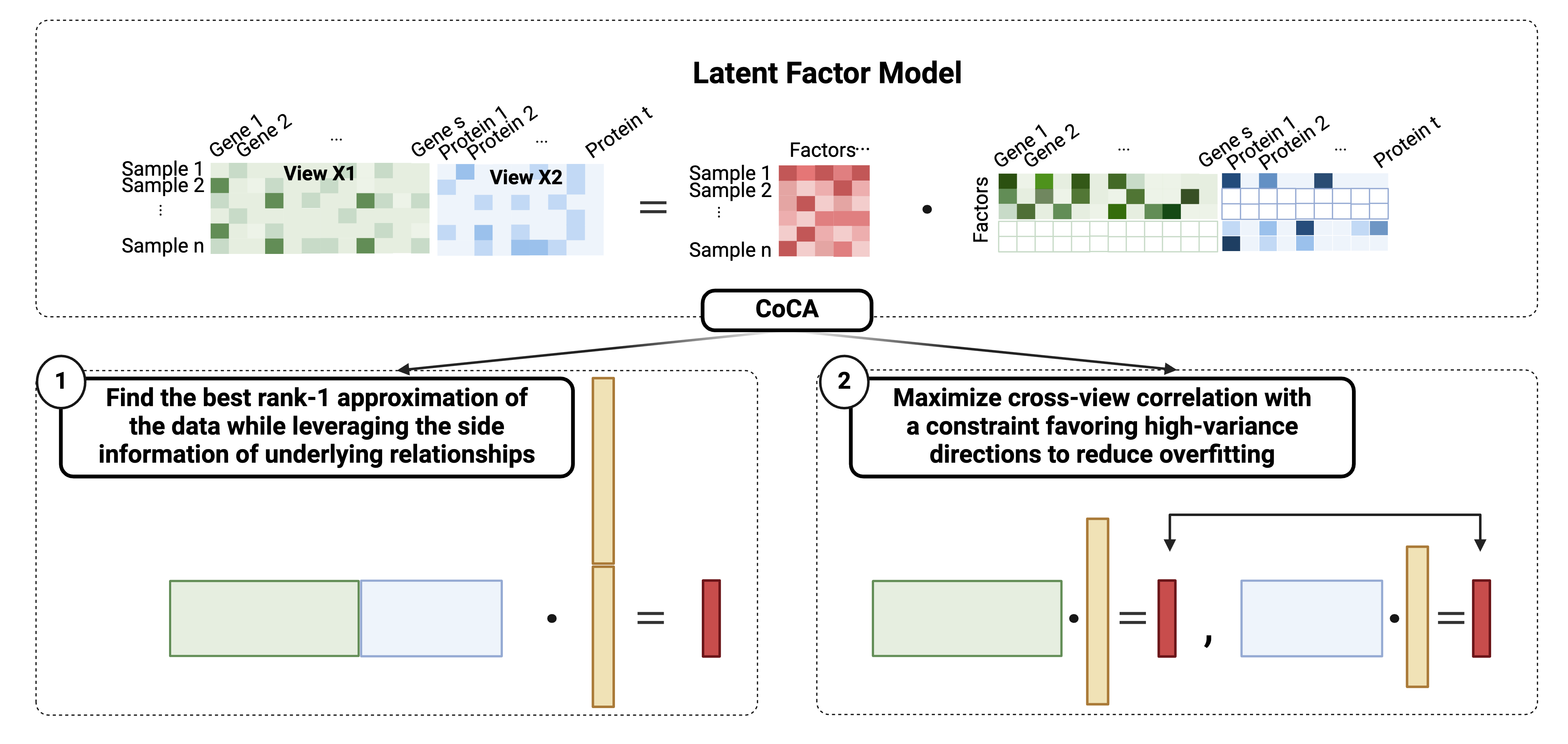}
    \caption{\emph{Intuition behind CoCA.} Consider a scenario where an important biological pathway, such as the p53 pathway involved in cell cycle regulation and tumor suppression, contains the most signal and drives the correlation between two data views (e.g., transcriptomics and proteomics). This pathway can be considered a latent factor, with key genes and corresponding proteins having loadings on this factor in both views. While there could be other pathways that explain large variance within each view, they may not be consistently shared between the views. CoCA can be understood from two perspectives. In the subpanel (1) on the bottom left, CoCA finds the best rank-1 approximation of the data by leveraging side information, i.e., shared underlying relationships between views, to improve approximation error. The assumption is that the best rank-1 approximation is the information that is redundantly expressed across views, and the agreement penalty exploits this structure to improve the approximation. Alternatively, in the subpanel (2) on the bottom right, CoCA can be seen as identifying the direction that maximizes correlation across views while incorporating a constraint favoring high-variance directions to reduce overfitting.}
    % under a latent factor model
    \label{fig:CoCA}
\end{figure}

% We will set the parameters in~\eqref{eqn:multi-view-latent-space-model} so that $\beta = (\beta_{1}, \beta_{2})$ is the leading population PC, and so that there are four sources of variation contributing to $x_{i,1}$ and $x_{i,2}$ in the following ways: (1) a latent variable $z$ that is shared across views and that has high variance-explained; (2) latent variables $s$ that are shared across views but have lower variance-explained; (3) latent variables $z_1,z_2$ that are view-specific; and the noise $\epsilon$. 

% Specifically, we consider the two views to have the same dimension $p_{1} = p_{2} = 4$ for simplicity and $n = 200$. 
% We set $\beta_{1} = \beta_{2} = \begin{pmatrix} 1 & 0 & 0 & 0 \end{pmatrix}^{\top}$, with $\beta_{1}$, $\bW_{1} \in \Reals^{p_1}$ and $\bB_{1} \in \Reals^{p_1}$ mutually orthogonal, and $\beta_{2}$, $\bW_{2} \in \Reals^{p_2}$ and $\bB_{2} \in \Reals^{p_2}$ mutually orthogonal; in addition, we set $\|\beta\|_2^2 > \|\bB\|_2^2$.
% The latent variables $z, s, z_{1}$, and $z_{2}$ are drawn from standard normal distributions, i.e., $z, w, z_{1}, z_{2} \sim N(0, 1)$.

\begin{figure}[h!]
    \centering
    %\begin{subfigure}{.45\textwidth}
        %\includegraphics[width=\textwidth]{figures/cooperative_svd/estimation_error.pdf}
    %\end{subfigure}
    %\begin{subfigure}{.45\textwidth}
    \includegraphics[width=0.8\textwidth]{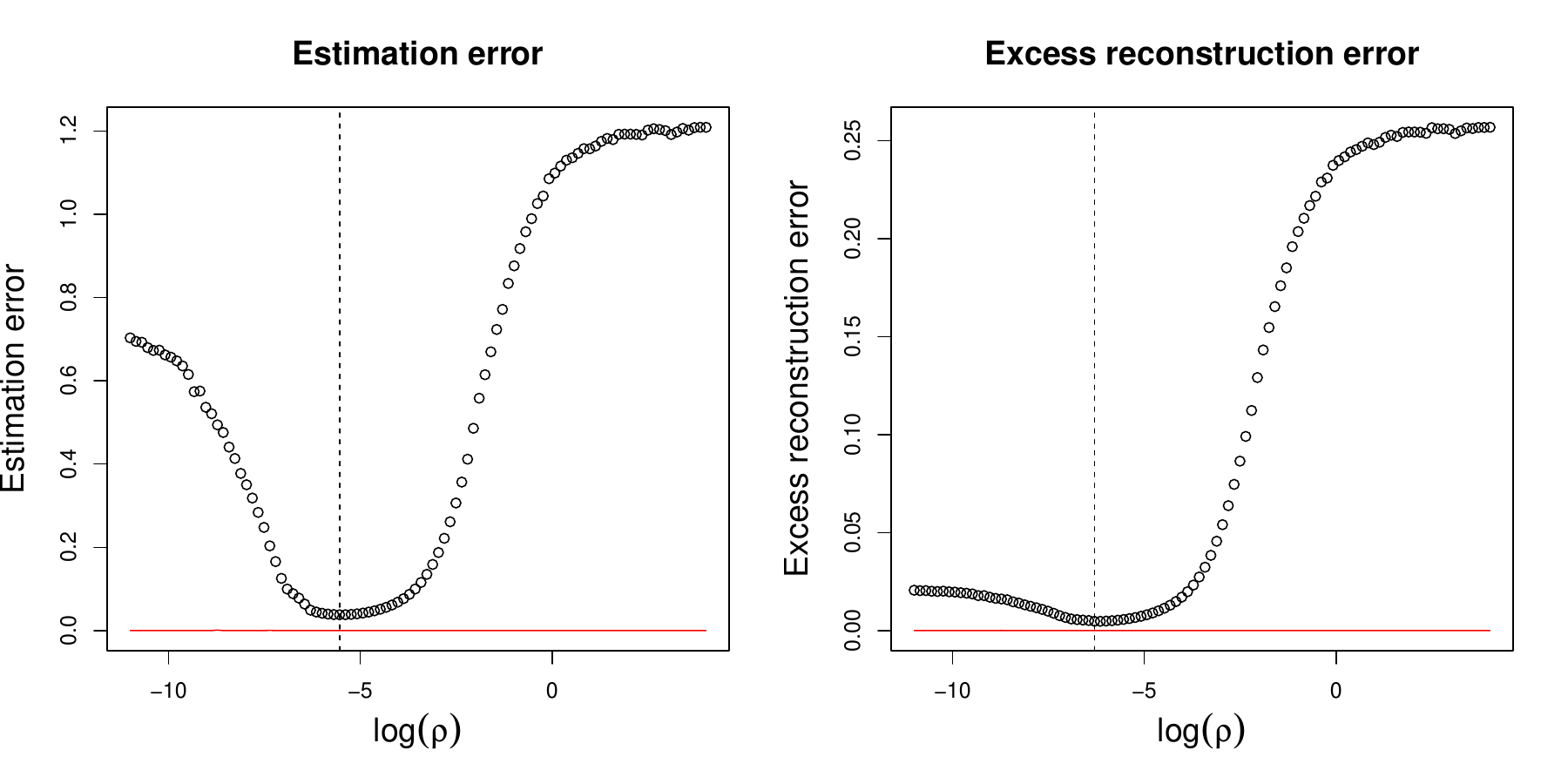}%{figures/cooperative_svd/illustrative_reconstruction.pdf}
    %\end{subfigure}
    \caption{\emph{An illustrative simulation study.} The plots show the estimation and excess reconstruction error of CoCA across different values of $\rho$ under a latent factor model. The red line corresponds to the error of the population CoCA solution across different values of $\rho$. %The reconstruction error measures how well the estimated component reconstructs the unseen test data. 
    Each point is an average over 100 Monte Carlo runs.} 
    %\ag{How can reconstruction error on a test set be 0?}}  
    \label{fig:cooperative-svd-factor-model}
\end{figure}

Figure \ref{fig:cooperative-svd-factor-model} visualizes the results when CoCA is computed on $n = 200$ samples drawn from this latent factor model. Two different criteria are used to evaluate performance of CoCA across different values of $\rho$. The left plot shows expected estimation error, which measures the difference between the estimated component $\hat{v}$ and the true component $\beta$:\footnote{We normalize $\beta \leftarrow \beta/\|\beta\|_2$ and $\hat{v} \leftarrow \hat{v}/\|\hat{v}\|_2$ before calculating the estimation and excess reconstruction error.} 
\begin{equation*}
    \E\Big[\min\{\|\hat{v} - \beta\|_2^2, \|-\hat{v} - \beta\|_2^2\}\Big]\footnote{The minimum is over $\hat{v},-\hat{v}$ because the CoCA solution is defined only up to $\pm 1$.}.
\end{equation*} 
The right plot shows the excess reconstruction error on a test set, where the test set $\bX_{\text{test}}$ is generated with the same set of parameters with more number of data points $n_{\text{test}} = 5000$.
The excess reconstruction error measures how well the estimated component $\hat{v}$ reconstructs the unseen test data as compared to the true component $\beta$, and is calculated as:
\begin{equation*}
    \E\Big[\frac{1}{n}\|\bX_{\text{test}} - \bX_{\text{test}} \hat{v} (\hat{v})^{\top}\|_F^2 - \frac{1}{n}\|\bX_{\text{test}} - \bX_{\text{test}} \beta (\beta)^{\top}\|_F^2\Big].
\end{equation*}
%\ag{Are you sure? The values of reconstruction error in the figure are very small. I think perhaps you want a $1/n$ in front?}

%Figure \ref{fig:cooperative-svd-factor-model} shows the reconstruction error on a test set, which measures how well the estimated component $\hat{v}$ reconstructs the unseen test data and is calculated as: \ag{Can you spell out a little more explicitly how reconstruction error is being calculated?}
% To enhance visualization, we present the logarithm of the reconstruction error. Lower values indicate better performance. 

Both the estimation error and excess reconstruction error plots reveal a clear U-shaped curve, with an optimal performance achieved at an intermediate value of $\rho$. 
The optimal point offers a marked improvement over the solution at either extreme of the solution path.
The red line in the plots corresponds to the solution of population CoCA, which remains at $\beta$ across different values of $\rho$. 
These results illustrate how CoCA effectively captures the shared signal and offers benefits over PCA and CCA in finite-sample scenarios.

In many applied problems with multiple data views, it can be reasonable to think of data as arising from a latent factor model in which there are (a few) shared and (many) individual latent factors. To give a concrete example in biomedicine, consider transcriptomics and proteomics data collected on a common set of cancer tissue samples (Figure \ref{fig:CoCA}). 
Suppose there is an important biological pathway, such as the p53 signaling pathway involved in cell cycle regulation and tumor suppression, that contains the most signal and makes the two views most correlated with each other. 
The pathway can be thought of as a latent factor: the key genes in the pathway (e.g. TP53, MDM2, CDKN1A) have expression levels in the transcriptomics data and corresponding protein abundances in the proteomics data that load strongly on this factor.
%The pathway can be thought of as a latent factor $z$, and the key genes in the pathway (e.g. TP53, MDM2, CDKN1A) have expression levels $\boldsymbol{x}^{(1)}$ and corresponding protein abundances $\boldsymbol{x}^{(2)}$ that load strongly on this factor via loadings $\boldsymbol{\beta}$. 
Besides this shared factor, there could be other pathways that explain large variance but are not shared consistently between both gene and protein expression, therefore unlikely to be the consistent biological signals we aim to identify. 
%For instance, the Notch signaling pathway, which plays a role in cell differentiation, might have high variability in gene expression in the transcriptomics data but not in protein abundances in the proteomics data.
%For instance, the Notch signaling pathway, which plays a role in cell differentiation, might have high variability in gene expression $\boldsymbol{x}^{(1)}$ but not in protein abundances $\boldsymbol{x}^{(2)}$. 
Alternatively, there might be other shared pathways, such as the Wnt signaling pathway involved in embryonic development, that are less relevant to the cancer phenotype of interest. 
Our goal is to specifically identify the important shared patterns, which could be most informative about the underlying biology. The simulation study of this section indicates CoCA can be an effective method for this kind of problem.

 % TODO: \ag{Not necessary pre-arXiv submission, but (imo) high on the priority list is fleshing this out with a more complete set of simulations, to put in an appendix.}
    
\section{Sparse cooperative component analysis (Sparse CoCA)}
\label{sec:sparse_CoCA}
\subsection{Sparse CoCA}

To encourage sparsity in the solution, we incorporate an $\ell_1$ penalty on $v$ into the objective in Problem~\eqref{eqn:cooperative-svd-2} and solve the following optimization problem: 
\begin{equation}
    \label{eqn:sparse-cooperative-svd}
    \min_{u,v} \|\bX -  u v^{\top}\|_{F}^2 + \rho \|\bX_1 v_1 - \bX_2 v_2\|_2^2 + \lambda \|v\|_1, \quad \st \|u\|_2^2 = 1. 
\end{equation}
We build upon Problem~\eqref{eqn:cooperative-svd-2} instead of Problem~\eqref{eqn:cooperative-svd} because the non-convex constraint on $v$ in the latter would make the optimization problem challenging to solve.
In Problem~\eqref{eqn:sparse-cooperative-svd}, $\rho \geq 0$ is the hyperparameter that controls the relative importance of the agreement penalty $\|\bX_1 v_1 - \bX_2 v_2\|_2^2$, and $\lambda \geq 0$ controls the sparsity of $v$ through the $\ell_1$ penalty.
We refer to this as \emph{sparse CoCA}. 
%\ag{Should be explicit that this is adding a penalty to~\eqref{eqn:cooperative-svd-2} not~\eqref{eqn:cooperative-svd}. And maybe explain why we are starting from~\eqref{eqn:cooperative-svd-2} not~\eqref{eqn:cooperative-svd}.}

%\subsection{Algorithm for Sparse CoCA}

%Problem~\eqref{eqn:sparse-cooperative-svd} is nonconvex.
We propose an alternating algorithm for optimizing the non-convex problem~\eqref{eqn:sparse-cooperative-svd}, inspired by the sparse PCA algorithm proposed by \cite{zou2006sparse}. 
%\ag{Should use the language "alternating algorithm" somewhere, I think.}
The algorithm iteratively optimizes over $v$ and $u$, fixing one and optimizing over the other. 
Specifically, the updates are as follows.
At step $k+1$, with a fixed value of $u^k$, we solve $v$ by solving the following minimization problem
\begin{equation}
\label{eqn:v-update}
\min_{v} \|\bX^{\top} u^k - v\|_2^2 + \rho \|\bX_1 v_1 - \bX_2 v_2\|_2^2 + \lambda \|v\|_1.
\end{equation}
The solution can be computed as follows. Letting
$$
\tilde{\bX} = \begin{bmatrix}
    \Identity_p \\
    \sqrt{\rho} \bX \bD 
\end{bmatrix},
\quad
\tilde{y} = \begin{bmatrix}
    \bX^{\top} u^k \\
    {\bf 0}_p
\end{bmatrix},
\quad \tilde{\beta} = \begin{bmatrix}
    v_1 \\
    v_2
\end{bmatrix},
$$
then Problem~\eqref{eqn:v-update} can be rewritten as:
%the equivalent problem with the same criteria to \eqref{eqn:v-update} is \ag{Equivalent in what sense?}
\begin{equation}
\label{eqn:v-update-lasso}
\min_{v}||\tilde y-\tilde \bX\tilde \beta||_2^2+\lambda \| \tilde \beta\|_1.
%(\|v_1\|_1 + \|v_2\|_1).
\end{equation}	
% HERE, need to change to 1/2
%\ag{Why not $\| \tilde \bbeta\|_1$?}
This is a form of the Lasso, and can be computed, for example
by the {\tt glmnet} package \citep{FHT2010}. 
%This new problem has $2n$ observations and $p$ features. \ag{Are you sure?}

With a fixed value of $v^{k + 1}$, the $u$-update is simply
			$$
			u^{k + 1} = \frac{\bX v^{k + 1}}{\|\bX v^{k + 1}\|_2}.
			$$

Let ${\rm Lasso}(\bX,{\vec{y}},\lambda)$ 
 denote the generic problem: %\ag{Need to be consistent about notation: $\min_{\beta}$ or ${\rm \min}_{\beta}$ or $\arg \min$, and $v_k$ or $v^k$ or $v^{k + 1}$.}
\be
{\underset{\beta}{\min} \; \|{\vec{y}}- \bX\beta\|_2^2 + \lambda \|\beta\|_1}.
\ee
We outline the alternating algorithm for sparse CoCA in Algorithm \ref{alg:iterative}.

\noindent
\begin{minipage}{.939\textwidth}
\begin{algorithm}[H]

\KwIn{Let ${\boldsymbol X}_1 \in \Reals^{n \times p_1}, {\boldsymbol X}_2 \in \Reals^{n \times p_2}$ --- representing two data views --- and ${\boldsymbol X} \in \Reals^{n \times p}$ be the concatenation of ${\boldsymbol X}_1$ and ${\boldsymbol X}_2$.
For fixed hyperparameters $\rho \geq 0$ and $\lambda \geq 0$.}

\vspace{2mm}

\For{$k \gets 0,1,2,\hdots$ {\normalfont until convergence}}{
At step $k+1$,
\begin{enumerate}
\item \emph{$v$-update:} At a fixed value of $u^k$, let %\ag{Again, bold vs. non-bold notation.}
$$
\tilde{\bX} = \begin{bmatrix}
    \Identity_p \\
    \sqrt{\rho} \bX \bD 
\end{bmatrix},
\quad
\tilde{y} = \begin{bmatrix}
    \bX^{\top} u^k \\
    {\bf 0}_p
\end{bmatrix}$$
    Update $v^{k+1}$ by solving ${\rm Lasso}(\tilde \bX, \tilde y, \lambda$).
    \vspace{0.3cm}
 \item \emph{$u$-update:} 
			$$
			u^{k + 1} = \frac{\bX v^{k + 1}}{\|\bX v^{k + 1}\|_2}.
			$$

\end{enumerate}
}
\caption{\em Alternating algorithm for sparse CoCA.}
\label{alg:iterative}
\end{algorithm}
\end{minipage}

The optimal value of $\rho$ and $\lambda$ can be determined using a validation set or cross-validation (CV) to estimate the test reconstruction error or other metrics based on the specific applications.
We describe the detailed CV procedures for CoCA in Appendix Section B.
%\ag{Again, do we explain exactly how to do this anywhere?}

{\bf Remark A.} Without sparsity, each $v$-update satisfies:
		$$
		v^{k + 1} = \frac{[\Identity + \rho \bD \bX^{\top} \bX \bD]^{-1} \bX^{\top} \bX v^k}{\|\bX v^k\|_2}.
		$$
These are the iterates of the power method, up to normalization, and it can be seen that the true solution $\hat{v}$ as defined in~\eqref{eqn:cooperative-svd-v} is a fixed point of the algorithm. 
Hence, convergence to the optimum without sparsity is guaranteed.
With sparsity, there is no longer a guarantee of convergence, but the algorithm behaves well in practice based on our experimental observations.

{\bf Remark B.}
We have explored alternative formulations of CoCA.
They correspond to different approaches to sparse PCA \citep{zou2018selective}. 
Without sparsity, they are equivalent to our final formulation, which corresponds to the SVD-based approach to sparse PCA. 
The key requirements for finalizing our formulation are two-fold: 
\begin{itemize}
    \item First, it encompasses PCA and CCA as special cases at the two ends of the solution path.
    \item Second, it is computationally tractable when sparsity is introduced, allowing the use of efficient algorithms like \texttt{glmnet} in an alternating optimization scheme. 
\end{itemize}
Our final formulation is the only one that satisfies these key requirements. We provide further details on the alternative formulations in Appendix Section \ref{sec:other_formulation}.

\subsection{Relation to other existing approaches}
We have mentioned in Section \ref{sec:pca_cca} the close connection of CoCA to PCA and CCA: setting $\rho=0$ or $\rho \to \infty$ in CoCA gives PCA and CCA, respectively, as special cases at the two ends of the solution path. 
%\ag{I would consider using the term ``solution path'' or ``path'' instead of spectrum, here and elsewheree.}

For high-dimensional settings, researchers have proposed various sparse variants of PCA \citep{zou2006sparse, jolliffe2003modified, d2004direct,  shen2008sparse, witten2009penalized, journee2010generalized, lu2012augmented, yuan2013truncated} and CCA \citep{parkhomenko2007genome, witten2009extensions, hardoon2011sparse, mai2019iterative, chen2019alternating, lindenbaum2021l0}, among others.
For example, Jolliffe et al. \cite{jolliffe2003modified} built upon the maximal variance characterization of PCA and proposed a method for sparse PCA called SCoTLASS to obtain sparse loadings. 
The first sparse principal component solves:
%by imposing an $\ell_1$ constraint on $v$.
%the loading vector $v_k$:%\ag{Check the language ``loading vector''.}
\begin{equation}
\label{eqn:scotlass}
%\max_{v} v_k^{\top} (\bX^{\top}\bX) v_k,
\max_{v} v^{\top} (\bX^{\top}\bX) v, \quad \st \|v\|^2_2 \leq 1, \|v\|_1 \leq c,
\end{equation}
where $c$ is a tuning parameter controlling the level of sparsity.
Subsequent components are obtained by solving the same problem with the additional constraint that they must be orthogonal to the previous components.
%subject to
%\begin{equation}
%\label{eqn:scotlass-constraints}
%v_k^{\top} v_k = 1 \quad \text{and} \quad \text{(for } k \geq 2) \quad v_h^{\top} v_k = 0, \quad h < k;
%\end{equation}
%and the extra constraints %\ag{The $\ell^2$ norms here and elsewhere should have subscripts, i.e. $\|\|_2$.}
%\begin{equation}
%\label{eqn:scotlass-l1}
%\sum_{j=1}^p \|v_{kj}\|_2 \leq t,
%\end{equation}
%for some tuning parameter $t$.
However, this direct formulation leads to a non-convex optimization problem that is computationally costly to solve.

After SCoTLASS, Zou et al. \cite{zou2006sparse} introduced a more computationally efficient SPCA algorithm for high-dimensional data. 
They showed that PCA can be formulated as a regression-type optimization problem, where sparse loadings are obtained by imposing the Lasso constraint on the regression coefficients $\beta$: 
%\ag{There is a way of writing~\eqref{eqn:spca-regression} using matrices, right? I would consider doing that, to make the similarities/differences to CoCA more apparent.}
\begin{equation}
\label{eqn:spca-regression}
%(\hat{\alpha}, \hat{\beta}) = \arg
\min_{\alpha,\beta} \|\bX - \bX\beta\alpha^{\top}\|_F^2 + \lambda_0\|\beta\|^2_2 + \lambda\|\beta\|_1, \quad \st \|\alpha\|_2^2 = 1,
%\min_{\alpha,\beta} \sum_{i=1}^n \|x_i - \alpha\beta^{\top} x_i\|^2 + \lambda\|\beta\|^2
\end{equation}
%subject to $\|\alpha\|^2 = 1$, and 
where $\lambda_0, \lambda \geq 0$, and $u \in \Reals^{p}, v \in \Reals^{p}$.
They leveraged the regression/reconstruction error property of PCA to derive sparse principal components and proposed an efficient alternating algorithm for solving Problem~\eqref{eqn:spca-regression}. 
%\ag{Should the norm on $\beta$ be $\ell_1$? Also the constraint on $\alpha$ should come inline, not afterwards.}

Furthermore, as PCA can also be solved via the SVD of the data matrix, there have also been methods proposed based on the SVD \citep{shen2008sparse, witten2009penalized}.
Specifically, Witten et al. \cite{witten2009penalized} proposed a penalized matrix decomposition (PMD) framework as follows: 
\begin{equation}
\label{eqn:pmd}
%\arg
\min_{d, u,v} \|\bX - duv^{\top}\|_F^2
\quad \st \|u\|_2^2 = 1, \|u\|_1 \leq c_1,
\|v\|_2^2 = 1, \|v\|_1 \leq c_2.
\end{equation}
An alternating algorithm was used to solve $u$ and $v$ iteratively.
In addition, they also showed that the PMD framework applied to a cross-product matrix solves penalized CCA.
We note that the approximation error term in the CoCA objective, $\|\bX - duv^{\top}\|_F^2$, is based on the best rank-$1$ approximation of the data matrix $\bX$ using the SVD, similar to the PMD framework of \cite{witten2009penalized}.
Furthermore, the computation of sparse CoCA draws inspiration from the algorithm proposed by Zou et al. \cite{zou2006sparse}, where we iteratively optimize for $u$ and $v$. 
Specifically, the update for $v$ can be reformulated as a Lasso problem, which can be efficiently solved using existing optimization techniques.

Another relevant approach is the \emph{joint and individual variation explained (JIVE)} method proposed by Lock et al. \cite{lock2013joint}. 
JIVE decomposes multiple data matrices into three terms: a low-rank approximation capturing joint structure between data views, low-rank approximations capturing patterns individual to each data view, and residual noise. 
This method allows for the simultaneous exploration of shared and view-specific patterns of variability in multi-view data.
Tang and Allen introduced another method to capture both individual and joint patterns across data views called \emph{integrated principal components analysis (iPCA)} \cite{tang2021integrated}. 
They employ a matrix-variate normal model and utilize penalized covariance estimators to extract these patterns.

CoCA also shares conceptual similarities with the supervised learning method \emph{cooperative learning} proposed by Ding et al. \cite{ding2022cooperative} for multi-view data. 
The method combines the usual squared error loss of predictions with an ``agreement'' penalty to encourage the predictions from different data views to agree. 
By varying the weight of the agreement penalty, 
cooperative learning encompasses the commonly-used early and late fusion and blended versions of these methods.
%cooperative learning encompasses a spectrum of solutions, including the commonly used early and late fusion approaches. 
In the regularized setting, considering feature matrices $\bX \in \mathbb{R}^{n \times p_x}$ and $\bZ \in \mathbb{R}^{n \times p_z}$, and target vector $y \in \mathbb{R}^{n}$, cooperative learning seeks to solve the following problem:
%find $\thx \in \mathbb{R}^{p_x}$ and $\thz \in \mathbb{R}^{p_z}$ that: %minimize:
\begin{equation}
%J(\thx, \thz) = 
\min_{\theta_x, \theta_z} \frac{1}{2} \|y - \bX\theta_x - \bZ\theta_z\|_2^2 + \frac{\rho}{2}\|(\bX\theta_x - \bZ\theta_z)\|_2^2 + \lambda_x P^x(\theta_x) + \lambda_z P^z(\theta_z),
\label{eq:obj1}
\end{equation}
where $\theta_x \in \mathbb{R}^{p_x}$ and $\theta_z \in \mathbb{R}^{p_z}$, $\rho$ controls the importance of the agreement term $\|(\bX\theta_x - \bZ\theta_z)\|_2^2$, and $P^x$ and $P^z$ are penalty functions.
Cooperative learning can be particularly effective when the different data views share some underlying relationship in their signals that can be exploited to boost the signal strength.
Furthermore, cooperative learning has also been extended to semi-supervised settings \citep{ding2023semi}, where the agreement penalty is utilized to leverage matched, unlabeled samples across data views to aid the learning process. 
CoCA further extends this concept of promoting agreement between views to unsupervised settings.

\subsection{Simulation study with sparse CoCA}

We evaluate sparse CoCA on simulated data drawn from the latent factor model with added noise as described in \eqref{eqn:multi-view-latent-space-model}. 
We introduce sparsity by letting some columns of $\bX_1$ and $\bX_2$ be pure noise dimensions.
For simplicity, we consider the two views to have the same dimension $p_{1} = p_{2}$, but this easily extends to views of different dimensions.

\begin{figure}[t!]
    %\centering
    %\begin{subfigure}{.45\textwidth}
%\includegraphics[width=\textwidth]{figures/sparse_CoCA/median_estimation_error_by_rho.pdf}
    %\end{subfigure}
    %\begin{subfigure}{.45\textwidth}
    \includegraphics[width=\textwidth]{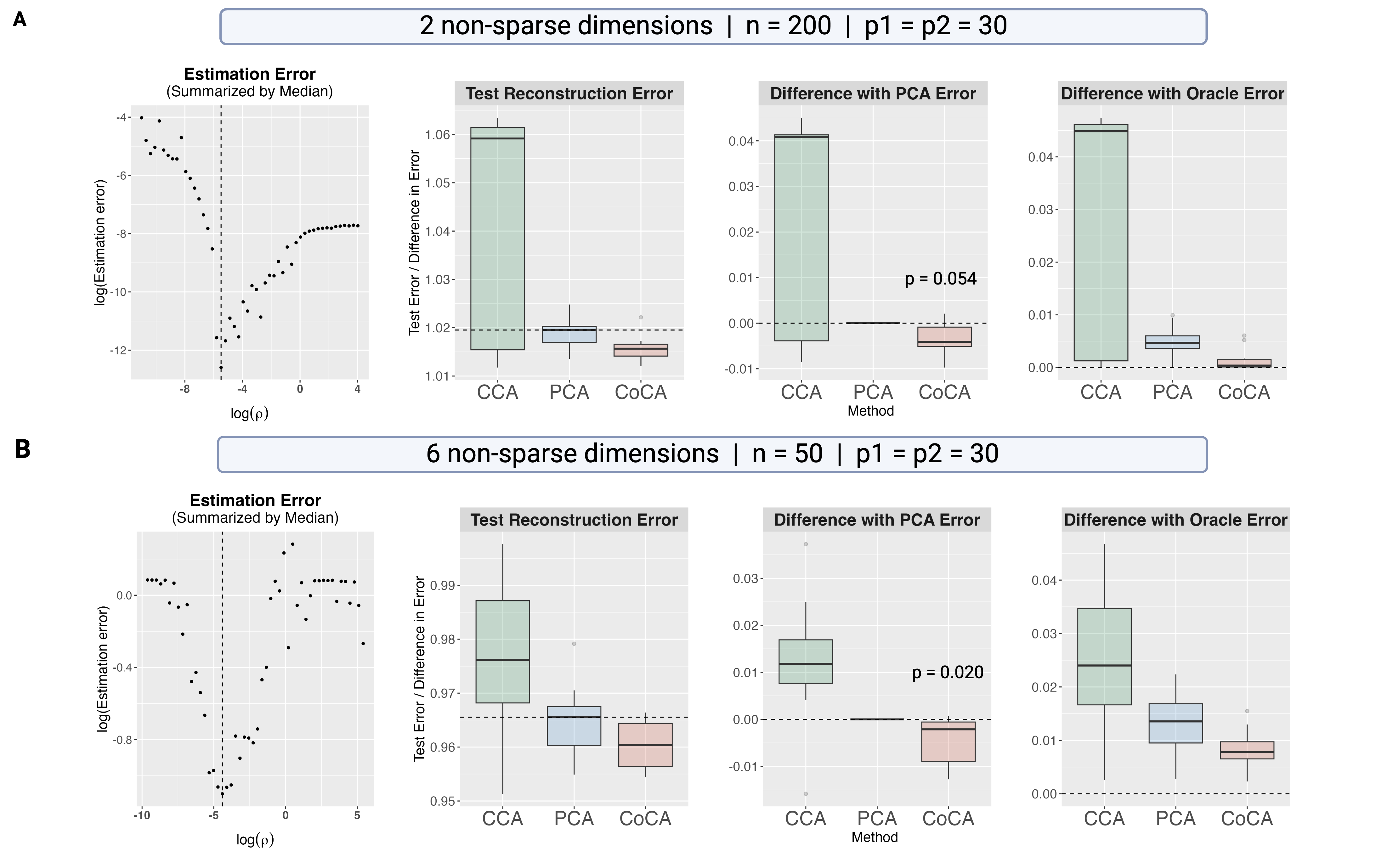}
    %\end{subfigure}
    \caption{\emph{Simulation results comparing CoCA with sparse PCA and CCA under different settings.} Panel A corresponds to the simulation setting with a simple true component with 2 non-sparse dimensions, $n = 200$, and $p = p_{1} = p_{2} = 30$. Panel B considers a denser component and a smaller training set with $n = 50$. In both settings, CoCA demonstrates benefits in reducing estimation and reconstruction error by leveraging shared underlying relationships between data views. The results are aggregated over 10 Monte Carlo runs.}
    \label{fig:sparse-cooperative-svd-factor-model}
\end{figure}

Figure \ref{fig:sparse-cooperative-svd-factor-model} presents the simulation results under the different settings. 
Panel A in Figure \ref{fig:sparse-cooperative-svd-factor-model} corresponds to the scenario where $\beta$ has 2 non-sparse dimensions.
%described above, where the true component to be estimated, i.e. the first column of $\bW$, is simple and has 2 non-sparse dimensions.
The training data consists of $n = 200$ data points with $p_{1} = p_{2} = 30$ features in each view.
A large test dataset with $5000$ data points is generated using the same set of parameters as the training set. 

The first subpanel of Panel A shows the estimation error across different values of $\rho$, aggregated over 10 Monte Carlo runs. 
The estimation error measures the difference between the true component used to generate the data and the estimated component. 
%Lower values indicate better performance, and the results are aggregated over 10 Monte Carlo runs.
The plot exhibits a U-shaped pattern, with the best performance achieved at an intermediate value of $\rho$. 
This optimal point demonstrates an estimation error over four orders of magnitude smaller than those seen at the two ends of the solution path.

%An intermediate value of $\rho$ achieves the best performance.
%\ag{A lot of the details above seem unnecessary, since we (will) have already described the latent factor model in Section 2.3. I would consider cutting redundant text, and standardizing the notation to match 2.3.}
Moreover, CoCA is compared with sparse PCA and CCA methods implemented using the \texttt{PMA} package \citep{witten2009penalized}.\footnote{Note that although CoCA with $\rho = 0$ and $\rho \to \infty$ corresponds to PCA and CCA, respectively, the \emph{sparse} PCA and CCA methods proposed in~\citep{witten2009penalized} are different than sparse CoCA with $\rho = 0$ and $\rho \to \infty$.} 
The optimal parameters for each method are selected using a validation set. 
The second subpanel shows the reconstruction error on the unseen test set.
Due to high variation across experiment runs, the results are also benchmarked against the sparse PCA method in each Monte Carlo run, with the difference shown in the third subpanel and a dotted horizontal line shown at zero. 
The last subpanel presents the difference with the oracle error in each run, where the oracle error is obtained by using the true component to reconstruct the test set.

CoCA outperforms PCA ($p = 0.054$) and CCA ($p = 0.007$) in reconstruction error, as determined by paired t-tests.
In addition, Panel B in Figure \ref{fig:sparse-cooperative-svd-factor-model} considers a more challenging setting with a denser component and a smaller training set of $n = 50$ data points. 
CoCA again demonstrates improved estimation error, and significantly outperforms PCA ($p = 0.020$) and CCA ($p = 0.002$) in reconstruction error.
%and reconstruction error, significantly improving upon PCA ($p = 0.020$) and CCA ($p = 0.002$).
%benefits in reducing estimation and reconstruction error. %by leveraging the shared underlying relationships between data views.
%Specifically, CoCA significantly improves upon PCA ($p = $) and CCA ($p = $), as determined by paired t-tests.

%\ag{I'm not sure I like the boxplots. They reflect Monte Carlo standard deviations -- not standard errors -- but I'm not sure that's a relevant quantity...and the size of the boxes kind of makes it look like all the methods are doing about the same. I assume you use the boxplots because if you did the dotplots everything looked extremely noisy? In that case I would suggest increasing the number of Monte Carlo runs.}

%\paragraph{PCA and sparse PCA}
%\paragraph{CCA and sparse CCA}
%\paragraph{Cooperative learning line of work}

%\subsection{Other formulations of CoCA}

% TO DO: details to be fixed
% 1. add encompassing PCA and CCA
% 2. add glmnet algorithm + mention the projection step can be problematic + check original beta constraint
% 3. add convex relaxation and algorithm

%Notice that even with $100$ Monte Carlo trials and a robust measure of average error (median), estimated error is somewhat unstable. Some insight into this phenomenon, and the reason for the difference between median and mean, can be gleaned from Figure~\ref{fig:sparse-cooperative-svd-factor-model-estimation-error}, which plots the distribution of estimation error.

\section{Real data examples}
\label{sec:real_data}

\subsection{Integration of CT scan-derived radiomics and laboratory measurements of COVID-19 patients}
We applied CoCA to integrate CT scan-derived radiomics features and laboratory result features measured on a cohort of 127 COVID-19 patients \citep{er2023multimodal}.
The goal of the analysis is to identify the component that captures important signals within data views while exhibiting strong correlations across data views.
In addition, we utilized the scores derived from the two data views for each patient under a linear discriminant analysis (LDA) model to predict their risk of severe disease progression requiring intensive care unit (ICU) admission. 
%projected component

The radiomics data consists of 1576 quantitative features extracted from CT scans, capturing various aspects of the imaged tissues, including density distribution and texture characteristics. These features provide an imaging-based characterization of the COVID-19-induced abnormalities. 
The clinical data, on the other hand, comprises 22 key laboratory measurements, including hemoglobin, white blood cell count, albumin, lactate dehydrogenase, D-dimer, C-reactive protein, and ferritin. These laboratory tests offer insights into the systemic effects of the disease and the patient's overall health status.

We split the dataset of 127 patients into training and test sets of 102 and 25 patients, respectively.
CV was employed to select the optimal hyperparameters for each method: $\rho$ and $\lambda$ for CoCA, and sparsity levels for sparse PCA and sparse CCA.
For sparse PCA and sparse CCA, we utilized the \texttt{PMA} package, which provides efficient implementations of these methods and allows for the specification of sparsity levels. 
The CV procedure involved deriving components using the training set and then estimating the test error by using the derived scores to predict ICU outcomes on the validation set.

\begin{figure}[t!]
    \centering
    \includegraphics[width=\textwidth]{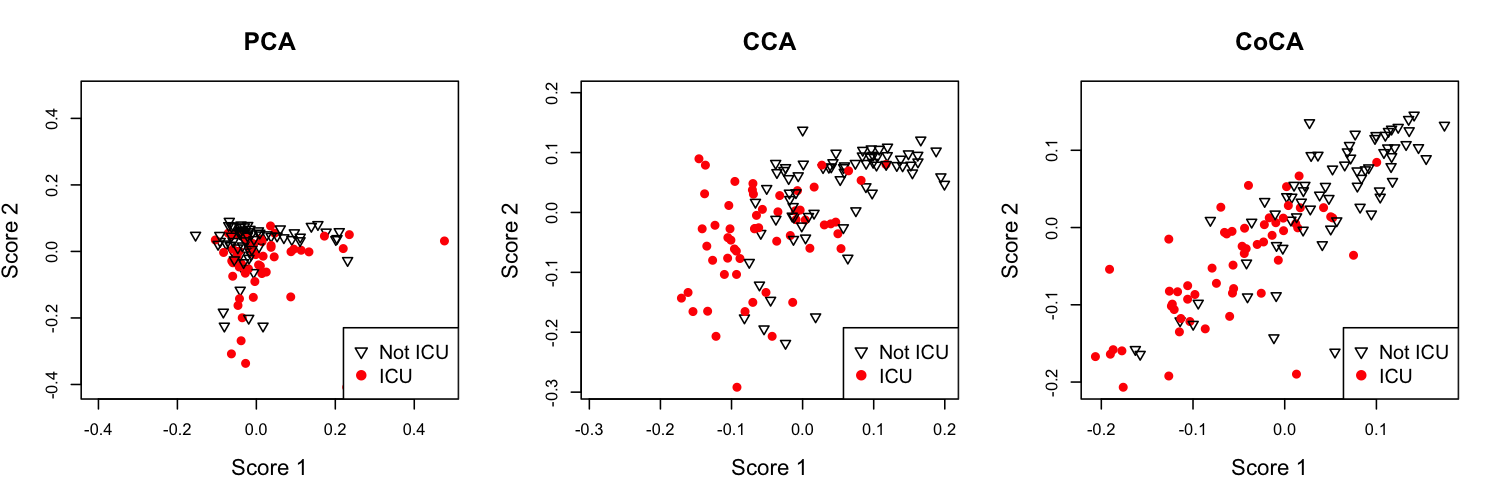}
    \caption{\emph{Comparison of multi-view scores derived from radiomics and clinical data views for a cohort of COVID-19 patients, using PCA, CCA, and CoCA, respectively.} We have seen this figure as a motivating example in the introduction. Here each point represents a patient, colored by ICU admission outcome. CoCA achieves clearer separation between ICU and non-ICU patients and better alignment between the two data views.}
    \label{fig:projections}
\end{figure}

Figure \ref{fig:projections} visualizes the scores for each patient obtained from sparse PCA, sparse CCA, and sparse CoCA. %projected component scores
Each point represents a patient, and the color indicates their ICU admission outcome. 
The plot provides a visual assessment of how well the derived components can separate patients with different disease progression outcomes. 
From the CoCA plot, we observe a clearer separation between patients who required ICU admission and those who did not, compared to the PCA and CCA plots. 
This suggests that the component learned by CoCA capture more informative and discriminative patterns that are associated with the severity of COVID-19 progression.
Moreover, the CoCA plot also exhibits better alignment of the scores derived from the two data modalities.
This indicates that CoCA effectively identifies components that are consistent and strongly correlated across the different data types, reflecting a shared underlying biological signal.

\begin{figure}[t!]
    \centering
    \begin{subfigure}{.39\textwidth}
        \includegraphics[width=\textwidth]{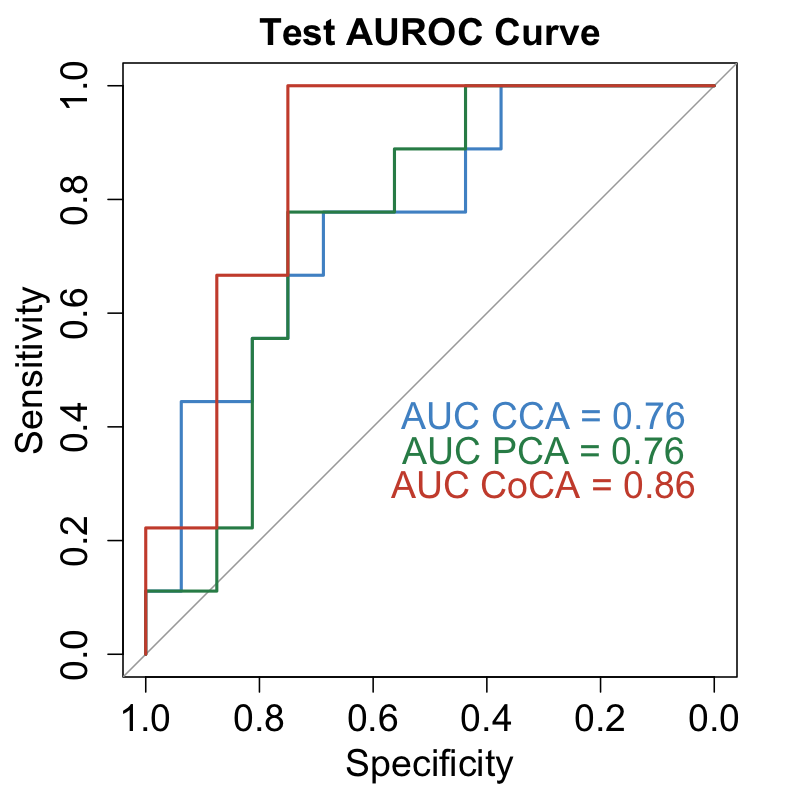}
    \end{subfigure}
    \begin{subfigure}{.6\textwidth}
        \includegraphics[width=\textwidth]{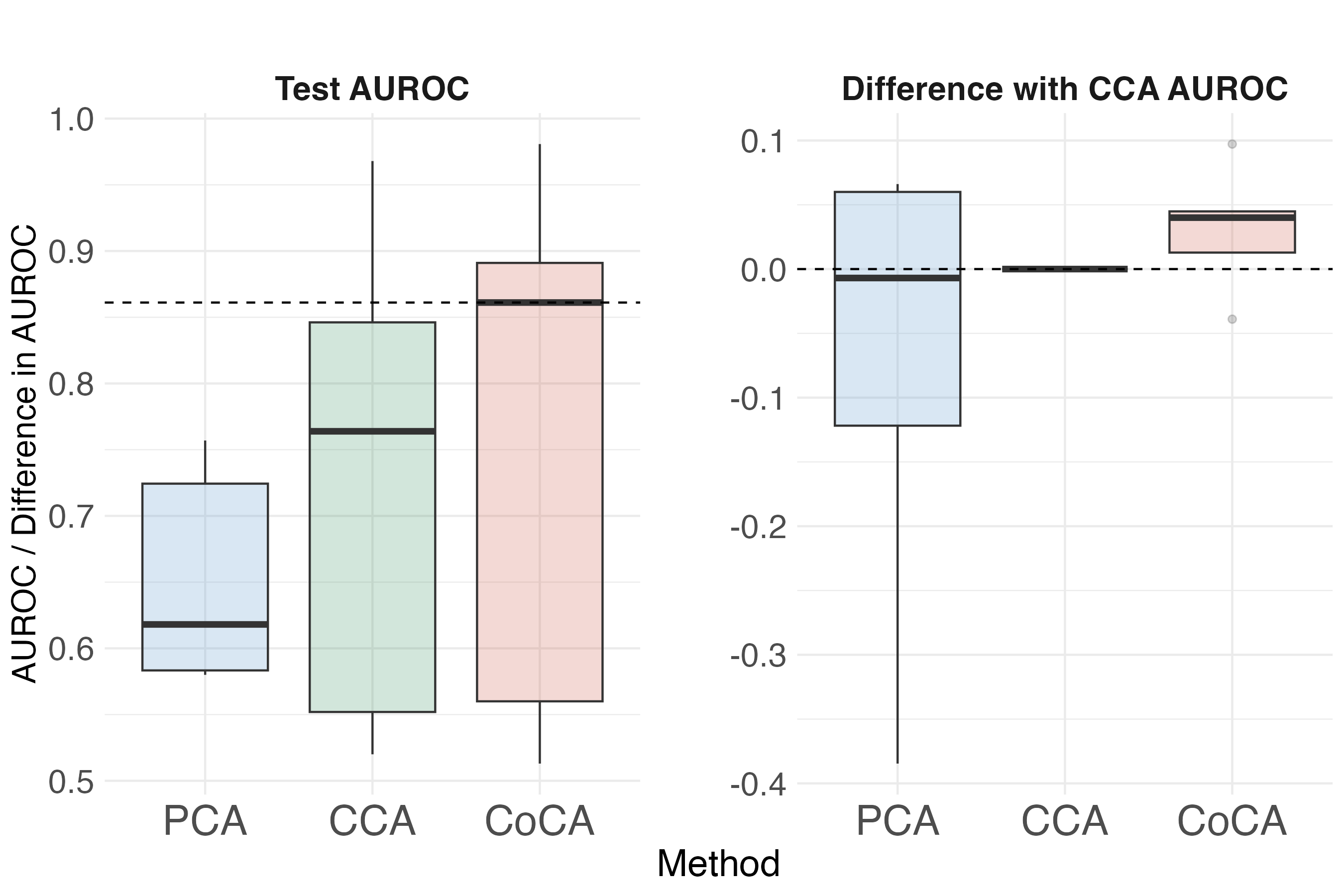}
    \end{subfigure}
    \caption{\emph{Test performance of using the multi-view scores from radiomics and clinical data for predicting ICU outcomes based on AUROC.} The left panel shows the AUROC curve on an unseen test set: CoCA outperforms both PCA and CCA, improving the AUROC from 0.76 to 0.86. The middle panel shows the AUROC performance across five experiment runs with different random splits of the training and test sets: the median AUROC for CoCA is higher than the upper quantile AUROC of both PCA and CCA. The right panel plots the difference in AUROC between each method and CCA across experiment runs, with positive values indicating that the method outperformes CCA.}
    \label{fig:auroc}
\end{figure}

In Figure \ref{fig:auroc}, we evaluate the test performance of using the scores for predicting ICU outcomes based on the area under the receiver operating characteristic curve (AUROC) metric. 
The left panel shows the AUROC curve on the test set: CoCA outperforms both PCA and CCA, improving the AUROC from 0.76 to 0.86. 
To assess the robustness of these findings, we conducted the experiment for five times with different random splits of the training and test sets.
The middle panel shows the AUROC performance across different experiment runs: the median AUROC for CoCA (represented by the center line) is higher than that of the upper quantile AUROC of both PCA and CCA. 
With the inherent variability in performance across different runs, we further benchmarked the methods against CCA in each experiment run. 
The right panel plots the difference in AUROC between each method and CCA across experiment runs, with positive values indicating that the method outperformes CCA. 

\begin{figure}[h!]
    \centering
    \begin{subfigure}{.39\textwidth}
        \includegraphics[width=\textwidth]{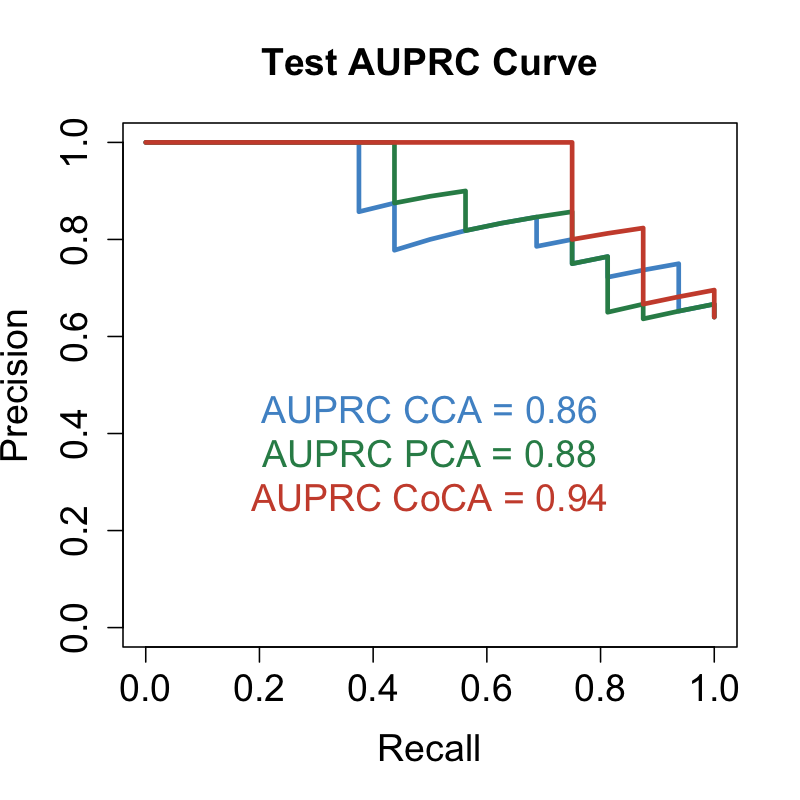}
    \end{subfigure}
    \begin{subfigure}{.6\textwidth}
        \includegraphics[width=\textwidth]{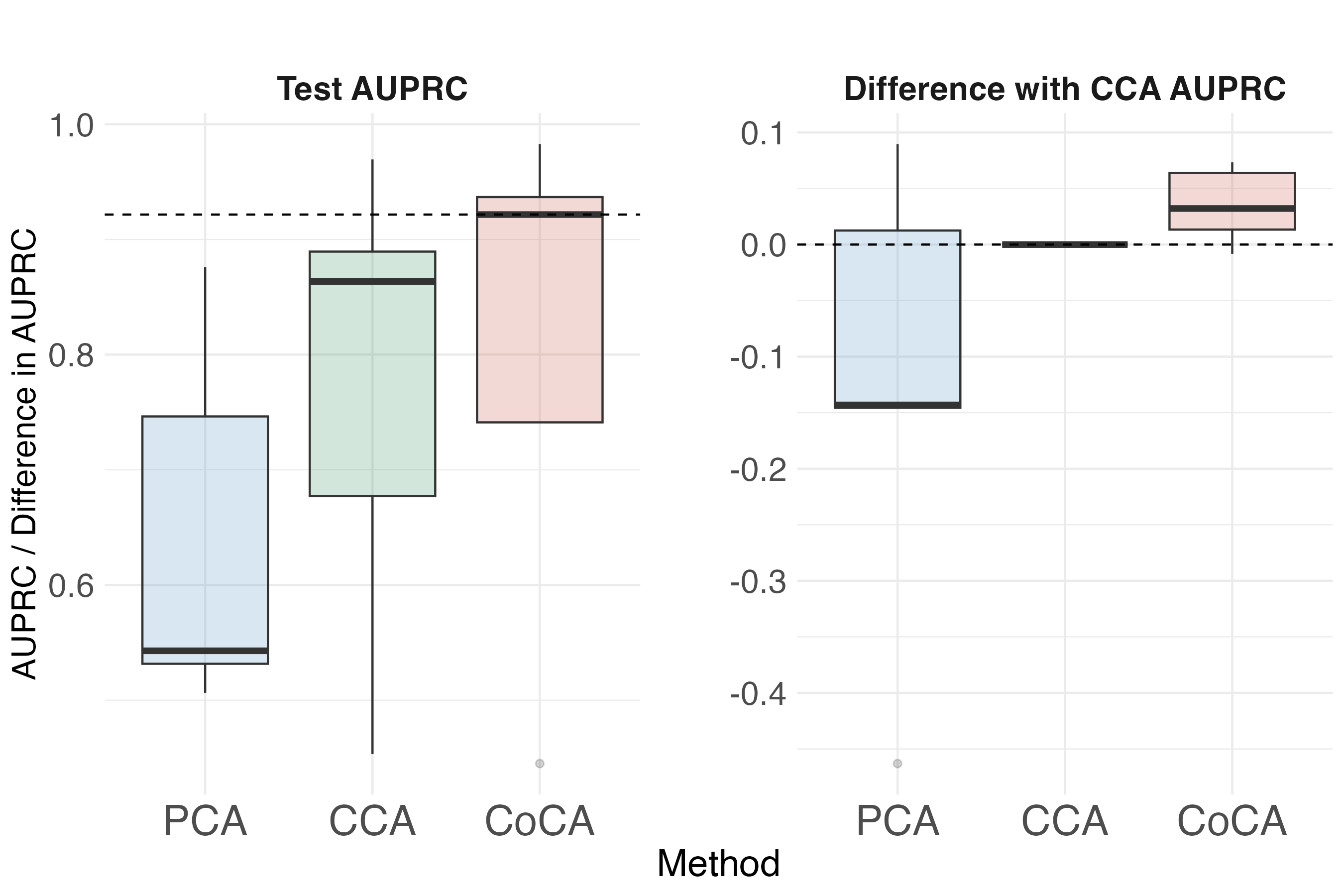}
    \end{subfigure}
    \caption{\emph{Test performance of using the multi-view scores from radiomics and clinical data for predicting ICU outcomes based on AUPRC.} The left panel shows the AUPRC curve on an unseen test set: CoCA outperforms both PCA and CCA, improving the AUPRC from 0.86 (CCA) and 0.88 (PCA) to 0.94 (CoCA). The middle panel shows the AUPRC performance across five experiment runs with different random splits of the training and test sets: the median AUPRC for CoCA is higher than the upper quantile AUPRC of both PCA and CCA. The right panel plots the difference in AUPRC between each method and CCA across experiment runs, with positive values indicating that the method outperformed CCA.}
    \label{fig:auprc}
\end{figure}

\iffalse
\begin{table}[h]
    \small
    \begin{center}
    \vspace{-1mm}
    \begin{tabular}{c|M{1.2cm}M{1.2cm}|M{1.2cm}M{1.2cm}||M{1.2cm}M{1.2cm}|M{1.2cm}M{1.2cm}}%{c|c{2.5cm}c{2.5cm}|cc|c}  %{l|ll|ll|l}
        \toprule
        \textbf{Methods} & \multicolumn{2}{c|}{\textbf{Test AUROC}} & \multicolumn{2}{c||}{\textbf{Relative to CCA}}  & \multicolumn{2}{c|}{\textbf{Test AUPRC}} & \multicolumn{2}{c}{\textbf{Relative to CCA}} \\
        & Mean & SD & Mean & SD & Mean & SD & Mean & SD  \\ \hline
        \midrule
        Sparse PCA & 0.653 & 0.083 & -0.077 & 0.188 & 0.641 & 0.163 & -0.130 & 0.212\\ 
        Sparse CCA & 0.730 & 0.192 & 0.000 & 0.000 & 0.771 & 0.207 & 0.000 & 0.000 \\
        \textbf{Sparse CoCA} & \textbf{0.761} & \textbf{0.210} & 0.031 & 0.050 & \textbf{0.805} & \textbf{0.221} & 0.035 & 0.034 \\
        \bottomrule
    \end{tabular}
    \end{center}
    \caption{{\em Test performance of Sparse PCA, Sparse CCA, and Sparse CoCA in predicting ICU outcomes}. The first two columns in the table show the mean and standard deviation (SD) of AUROC on the test set across different splits of the training and test sets. The third and fourth column show the AUROC difference relative to sparse CCA, with positive values indicating better performance than Sparse CCA.}
    \label{tab:results_covid}
\end{table}
\fi

Moreover, to account for the class imbalance in the data, where 57 out of 127 patients required ICU admission, we also evaluated the performance using the area under the precision-recall curve (AUPRC). 
As before, the left panel of Figure \ref{fig:auprc} shows the AUPRC curve on the test set for one split of the training and test sets. 
The middle panel shows the AUPRC performance for each method across five different random splits of the training and test sets. 
The right panel plots the difference in AUPRC between each method and CCA. 
These results highlight the effectiveness of CoCA.
By leveraging the shared information across the radiomics and clinical data views, CoCA is able to extract more informative components that are consistently present across views and more predictive of disease progression.

We also performed model interpretation by examining the top selected features from the two data modalities. 
The selected radiomics features include the first-order statistics such as total energy in the LLL (low-low-low) wavelet decomposed image, the gray level dependence matrix (GLDM) dependence variance in the LLH (low-low-high) wavelet decomposed image, and the gray level size zone matrix (GLSZM) small area emphasis in the LLL (low-low-low) wavelet decomposed image, among others. 
They capture the intensity and textural characteristics of the imaged tissues, reflecting the heterogeneity and complexity of COVID-19-induced abnormalities in the lungs. 
On the other hand, the top selected clinical features, albumin (ALB) and lactate dehydrogenase (LDH), are associated with disease severity and tissue damage. 
Specifically, low ALB levels indicate a severe systemic inflammatory response, potentially leading to pulmonary edema, while high LDH levels mark the extent of lung injury. 
The correlation between these radiomics and clinical features can be explained by the underlying pathophysiology of COVID-19, where the systemic inflammatory response and tissue damage manifest as alterations in the intensity and texture of the CT images. 
CoCA's integration of these features allows for a more comprehensive characterization of disease severity and progression, potentially improving risk stratification and guiding clinical decision-making.

%The potential correlation between these radiomics and clinical features can be explained by the underlying pathophysiology of COVID-19. Low albumin levels are indicative of a severe systemic inflammatory response, which can lead to increased vascular permeability and pulmonary edema. These changes in the lung parenchyma may manifest as alterations in the intensity and texture of the CT images, captured by features such as wavelet LLL firstorder TotalEnergy and wavelet LLL glszm SmallAreaEmphasis. On the other hand, high lactate dehydrogenase levels are a marker of tissue damage and cell death, reflecting the extent of lung injury caused by the viral infection. The degree of tissue damage may be reflected in the complexity and heterogeneity of the lung texture, as measured by features like wavelet LLH gldm DependenceVariance.

\subsection{Integrating epithelial and stromal gene expression data of breast ductal carcinoma in situ (DCIS) patients}

We also applied CoCA to integrate epithelial and stromal gene expression data from a study on ductal carcinoma in situ (DCIS), the most common precursor of invasive breast cancer (IBC) \cite{strand2022molecular}.
In the dataset, the Resource of Archival Breast
Tissue (RAHBT) cohort contained 78 patients, including 17 patients with concurrent contralateral IBC, 27 patients with disease recurrence, and 34 controls without disease recurrence.
The data consisted of paired epithelial and stromal gene expression profiles obtained through laser capture microdissection, allowing for the separate analysis of these two key tissue compartments. 
The goal of the analysis was to identify the component that captures important signals within and across different molecular data views, providing insights into the spectrum of molecular changes in DCIS and potential predictors of disease progression. 
In addition, we utilized the scores derived from the two data views for each patient under a LDA model to classify DCIS patients into concurrent IBC, disease recurrence and no-recurrence. 

Both epithelial and stromal gene expression data consist of 60,662 gene expression features, which were screened by their variance across the subjects.
We split the dataset of 78 patients into training and test sets of 70 and 8 patients, respectively. 
To ensure robustness of our results, we conducted the same set of experiments across 10 different random splits of the training and test sets.
CV was employed to select the optimal hyperparameters for each method: $\rho$ and $\lambda$ for CoCA, and sparsity levels for sparse PCA and sparse CCA.
For sparse PCA and sparse CCA, we utilized the \texttt{PMA} package. 
The CV procedure involved deriving components using the training set and then estimating the test error by using the scores to predict disease outcomes on the validation set.

\begin{figure}[t!]
    \centering
    \includegraphics[width=\textwidth]{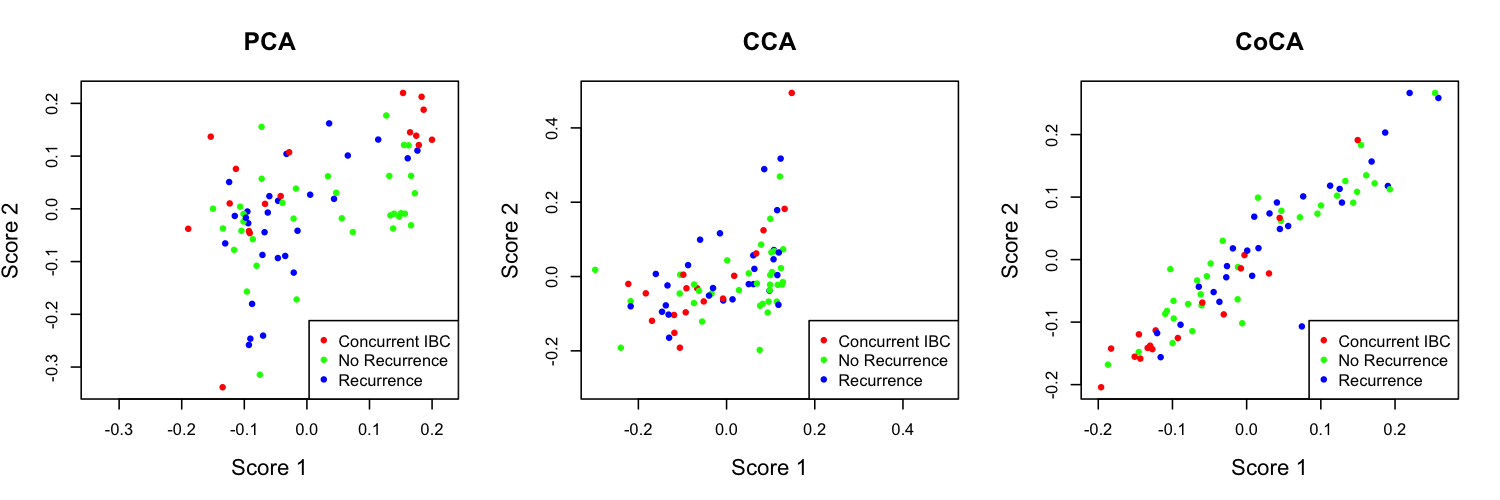}
    \caption{\emph{Comparison of multi-view scores derived from epithelial and stromal gene expression data for a cohort of DCIS patients with different disease outcomes, using PCA, CCA, and CoCA, respectively.}
    %Comparison of multi-view scores based on the components obtained from PCA, CCA, and CoCA for DCIS patients with different disease outcomes, using epithelial and stromal gene expression data.
    Here each point represents a patient, colored by differnt outcome groups, i.e. concurrent contralateral IBC, disease recurrence and no-recurrence. CoCA achieves clearer separation between patients with different outcomes and better alignment between the two molecular data views.}
    \label{fig:projections_dcis}
\end{figure}

Figure \ref{fig:projections_dcis} visualizes the multi-view scores for each patient obtained from sparse PCA, sparse CCA, and sparse CoCA. 
Each point represents a patient, and the color indicates their disease outcome, i.e. concurrent contralateral IBC, future recurrence, or no recurrence.
The plot provides a visual assessment of how well the derived components can separate patients with different disease outcomes. 
The CoCA plot demonstrates a more distinct separation of patients with differnt outcomes, as well as enhanced alignment between the two sets of scores derived from epithelial and stromal data views.
This suggests that CoCA more effectively captures shared biological signals across tissue compartments that may be crucial in distinguishing DCIS progression patterns and risk of invasive cancer development.

%{\color{blue}To do for Daisy: finish this section + add figures}
Table \ref{tab:results_cancer} compares the performance of Sparse PCA, Sparse CCA, and Sparse CoCA in differentiating DCIS patients into outcome groups. 
The multi-class AUROC metric \cite{hand2001simple} is used for evaluation. 
The table shows the mean and standard deviation (SD) of both the absolute AUROC values and the performance relative to CCA across different splits of the training and test sets.
%multiple experiment runs.
Sparse CoCA achieves the highest mean AUROC (0.753) on unseen test sets and shows the best performance relative to CCA (+0.060), suggesting that it offers superior discriminative power for classifying DCIS outcomes.

\begin{table}[h]
    %\tiny
    \begin{center}
    %\vspace{-1mm}
    \begin{tabular}{c|M{1.2cm}M{1.2cm}|M{1.2cm}M{1.2cm}}%{c|c{2.5cm}c{2.5cm}|cc|c}  %{l|ll|ll|l}
        \toprule
        \textbf{Methods} & \multicolumn{2}{c|}{\textbf{Test AUROC}} & \multicolumn{2}{c}{\textbf{Relative to CCA}} \\
        & Mean & SD & Mean & SD \\ \hline
        \midrule
        Sparse PCA & 0.628 & 0.141 & -0.066 & 0.010 \\ 
        Sparse CCA & 0.694 & 0.110 & 0.000 & 0.000 \\
        \textbf{Sparse CoCA} & \textbf{0.753} & \textbf{0.089} & \textbf{0.060} & \textbf{0.043} \\
        \bottomrule
    \end{tabular}
    \end{center}
    \caption{{\em Test performance of Sparse PCA, Sparse CCA, and Sparse CoCA in differentiating DCIS patients}. The first two columns in the table show the mean and standard deviation (SD) of the multi-class AUROC on the test set across different splits of the training and test sets. The third and fourth column show the AUROC difference relative to sparse CCA, with positive values indicating better performance than sparse CCA.}
    \label{tab:results_cancer}
\end{table}

We also performed model interpretation by examining the top selected features from the paired epithelial and stromal gene expression profiles.
Several genes were consistently selected as important features in both epithelial and stromal compartments, including IGKC, TFF1, MUCL1, SLC39A6, and ANKRD30A. 
IGKC, a key component of the humoral immune system, has been associated with improved distant metastasis-free survival in early breast cancer, suggesting a potential protective role of tumor-infiltrating immune cells \cite{schmidt2021prognostic,schmidt2012immunoglobulin}. 
TFF1, an estrogen-regulated protein, has been shown to stimulate migration of human breast cancer cells, potentially contributing to disease recurrence \cite{prest2002estrogen}. These genes are worth further investigation for their roles in DCIS progression.

\section{Discussion}
\label{sec:discussion}

In this paper, we introduce Cooperative Component Analysis (CoCA), a novel unsupervised learning method for integrating multiple data views, by identifying the component that simultaneously captures significant within-view variance and exhibits strong cross-view correlation. 
CoCA encourages alignment across data views through an agreement penalty.
By varying the weight of the agreement penalty in the objective, CoCA encompasses the commonly used principal component analysis (PCA) and canonical correlation analysis (CCA) as special cases at the two ends of the solution path.
%we obtain a spectrum of solutions that include the commonly used principal component analysis (PCA) and canonical correlation analysis (CCA) as special cases.
This allows CoCA to choose the degree of agreement in a data-adaptive manner.

Additionally, this approach has promises to be extended to scenarios where explicit grouping information is unavailable or suboptimal. 
In such cases, we adaptively identify the optimal grouping that best captures the underlying shared structure across variables, with the potential to improve upon PCA without prior knowledge of the grouping.
Moreover, it is also possible to define a rank-$K$ generalization of CoCA, by replacing $u$ and $v$ with rank-$K$ orthogonal matrices in~\eqref{eqn:cooperative-svd}. However, it is more difficult to define a computationally tractable rank-$K$ sparse formulation of CoCA, due to challenges introduced by the orthogonality constraints, and so we leave this for future work. 

Furthermore, to enhance interpretability, CoCA incorporates the Lasso penalty to yield a sparse component. 
This facilitates the identification of key features driving the observed patterns. 
The effectiveness of CoCA has implications for improving our understanding of complex systems and uncovering novel insights in an era of increasingly multi-view data.
%with increasingly more multi-view data.
%Moreover, it is possible to extend CoCA to identify rank-$K$ co
% Alden's effort: 

\section*{Acknowledgements}
We would like to thank Andreas Buja, Trevor Hastie, and Jonathan Taylor for helpful discussions.
D.Y.D was supported by the Stanford Graduate Fellowship (SGF) and Stanford Data Science Scholarship.
M.W.S was supported by National Institute of Health (NIH) T15 NLM007033.
 R.T. was supported by the National Institutes of Health (5R01EB001988-16) and the National Science Foundation (19DMS1208164).

%adaptively balancing the trade-off between explaining within-view variance and maximizing cross-view correlation. 
%It encourages alignment across data views while allowing for flexibility in the degree of alignment based on the underlying structure of the data.

%CoCA offers several advantages over existing approaches:
%\begin{enumerate}
    %\item It identifies the component that simultaneously captures significant within-view variance and exhibits strong cross-view correlation.
    %\vspace{-0.3cm}
    %\item By varying the weight of the agreement penalty, it encompasses both PCA and CCA as special cases. CoCA adaptively balances the trade-off between explaining within-view variance and maximizing cross-view correlation based on the data's inherent structure.
    %\vspace{-0.3cm}
    %\item By incorporating sparsity-inducing penalties, CoCA yields a sparse and interpretable component and facilitates the identification of key features driving the observed patterns.
%\end{enumerate}

%Other approaches, other objectives in the main text, algorithm in the appendix
%Simulation, weak signals
%Roadmap, to do list (done)

%\newpage
\bibliography{reference} 

\begin{thebibliography}{10}

\bibitem{pearson1901liii}
Karl Pearson.
\newblock On lines and planes of closest fit to systems of points in space.
\newblock {\em The London, Edinburgh, and Dublin Philosophical Magazine and
  Journal of Science}, 2(11):559--572, 1901.

\bibitem{reich2008principal}
David Reich, Alkes~L Price, and Nick Patterson.
\newblock Principal component analysis of genetic data.
\newblock {\em Nature Genetics}, 40(5):491--492, 2008.

\bibitem{alter2000singular}
Orly Alter, Patrick~O Brown, and David Botstein.
\newblock Singular value decomposition for genome-wide expression data
  processing and modeling.
\newblock {\em Proceedings of the National Academy of Sciences},
  97(18):10101--10106, 2000.

\bibitem{greenacre2022principal}
Michael Greenacre, Patrick~JF Groenen, Trevor Hastie, Alfonso~Iodice d’Enza,
  Angelos Markos, and Elena Tuzhilina.
\newblock Principal component analysis.
\newblock {\em Nature Reviews Methods Primers}, 2(1):100, 2022.

\bibitem{novembre2008interpreting}
John Novembre and Matthew Stephens.
\newblock Interpreting principal component analyses of spatial population
  genetic variation.
\newblock {\em Nature Genetics}, 40(5):646--649, 2008.

\bibitem{allee2022pca}
Kristian~D. Allee, Chuong Do, and Fellipe~G. Raymundo.
\newblock Principal component analysis and factor analysis in accounting
  research.
\newblock {\em Journal of Financial Reporting}, 7:1–39, September 2022.

\bibitem{guellis2020pca}
Cristiane Guellis, Daniele~C. Valério, Guilherme~G. Bessegato, Marcela
  Boroski, Josiane~C. Dragunski, and Cleber~A. Lindino.
\newblock Non-targeted method to detect honey adulteration: Combination of
  electrochemical and spectrophotometric responses with principal component
  analysis.
\newblock {\em Journal of Food Composition and Analysis}, 89:103466, 2020.

\bibitem{ghorbani2020pca}
Mahsa Ghorbani and Edwin K.~P. Chong.
\newblock Stock price prediction using principal components.
\newblock {\em Plos One}, 15:e0230124, March 2020.

\bibitem{cunningham2014dimensionality}
John~P Cunningham and Byron~M Yu.
\newblock Dimensionality reduction for large-scale neural recordings.
\newblock {\em Nature neuroscience}, 17(11):1500--1509, 2014.

\bibitem{hotelling1992relations}
Harold Hotelling.
\newblock Relations between two sets of variates.
\newblock In {\em Breakthroughs in Statistics: Methodology and Distribution},
  pages 162--190. Springer, 1992.

\bibitem{witten2009penalized}
Daniela~M Witten, Robert Tibshirani, and Trevor Hastie.
\newblock A penalized matrix decomposition, with applications to sparse
  principal components and canonical correlation analysis.
\newblock {\em Biostatistics}, 10(3):515--534, 2009.

\bibitem{gao2017sparse}
Chao Gao, Zongming Ma, and Harrison~H Zhou.
\newblock Sparse cca: Adaptive estimation and computational barriers.
\newblock {\em The Annals of Statistics}, pages 2074--2101, 2017.

\bibitem{witten2009extensions}
Daniela~M Witten and Robert~J Tibshirani.
\newblock Extensions of sparse canonical correlation analysis with applications
  to genomic data.
\newblock {\em Statistical Applications in Genetics and Molecular Biology},
  8(1), 2009.

\bibitem{janse2021cca}
Roemer~J Janse, Tiny Hoekstra, Kitty~J Jager, Carmine Zoccali, Giovanni
  Tripepi, Friedo~W Dekker, and Merel Van~Diepen.
\newblock Conducting correlation analysis: important limitations and pitfalls.
\newblock {\em Clinical Kidney Journal}, 14(11):2332–2337, 2021.

\bibitem{er2023multimodal}
Ahmet~Gorkem Er, Daisy~Yi Ding, Berrin Er, Mertcan Uzun, Mehmet Cakmak,
  Christoph Sadee, Gamze Durhan, Mustafa~Nasuh Ozmen, Mine~Durusu Tanriover,
  Arzu Topeli, et~al.
\newblock Multimodal data fusion using sparse canonical correlation analysis
  and cooperative learning: a covid-19 cohort study.
\newblock {\em NPJ Digital Medicine}, 7(1):117, 2024.

\bibitem{eckart1936approximation}
Carl Eckart and Gale Young.
\newblock The approximation of one matrix by another of lower rank.
\newblock {\em Psychometrika}, 1(3):211--218, 1936.

\bibitem{johnstone2001distribution}
Iain~M Johnstone.
\newblock On the distribution of the largest eigenvalue in principal components
  analysis.
\newblock {\em The Annals of Statistics}, 29(2):295--327, 2001.

\bibitem{tipping1999probabilistic}
Michael~E Tipping and Christopher~M Bishop.
\newblock Probabilistic principal component analysis.
\newblock {\em Journal of the Royal Statistical Society Series B: Statistical
  Methodology}, 61(3):611--622, 1999.

\bibitem{paul2007asymptotics}
Debashis Paul.
\newblock Asymptotics of sample eigenstructure for a large dimensional spiked
  covariance model.
\newblock {\em Statistica Sinica}, pages 1617--1642, 2007.

\bibitem{johnstone2009consistency}
Iain~M Johnstone and Arthur~Yu Lu.
\newblock On consistency and sparsity for principal components analysis in high
  dimensions.
\newblock {\em Journal of the American Statistical Association},
  104(486):682--693, 2009.

\bibitem{bach2005probabilistic}
Francis~R Bach and Michael~I Jordan.
\newblock A probabilistic interpretation of canonical correlation analysis.
\newblock 2005.

\bibitem{chen2013sparse}
Mengjie Chen, Chao Gao, Zhao Ren, and Harrison~H Zhou.
\newblock Sparse cca via precision adjusted iterative thresholding.
\newblock {\em arXiv preprint arXiv:1311.6186}, 2013.

\bibitem{zou2006sparse}
Hui Zou, Trevor Hastie, and Robert Tibshirani.
\newblock Sparse principal component analysis.
\newblock {\em Journal of Computational and Graphical Statistics},
  15(2):265--286, 2006.

\bibitem{FHT2010}
J.~Friedman, T.~Hastie, and R.~Tibshirani.
\newblock Regularization paths for generalized linear models via coordinate
  descent.
\newblock {\em Journal of Statistical Software}, 33:1--22, 2010.

\bibitem{zou2018selective}
Hui Zou and Lingzhou Xue.
\newblock A selective overview of sparse principal component analysis.
\newblock {\em Proceedings of the IEEE}, 106(8):1311--1320, 2018.

\bibitem{jolliffe2003modified}
Ian~T Jolliffe, Nickolay~T Trendafilov, and Mudassir Uddin.
\newblock A modified principal component technique based on the lasso.
\newblock {\em Journal of Computational and Graphical Statistics},
  12(3):531--547, 2003.

\bibitem{d2004direct}
Alexandre d'Aspremont, Laurent Ghaoui, Michael Jordan, and Gert Lanckriet.
\newblock A direct formulation for sparse pca using semidefinite programming.
\newblock {\em Advances in Neural Information Processing Systems}, 17, 2004.

\bibitem{shen2008sparse}
Haipeng Shen and Jianhua~Z Huang.
\newblock Sparse principal component analysis via regularized low rank matrix
  approximation.
\newblock {\em Journal of Multivariate Analysis}, 99(6):1015--1034, 2008.

\bibitem{journee2010generalized}
Michel Journ{\'e}e, Yurii Nesterov, Peter Richt{\'a}rik, and Rodolphe
  Sepulchre.
\newblock Generalized power method for sparse principal component analysis.
\newblock {\em Journal of Machine Learning Research}, 11(2), 2010.

\bibitem{lu2012augmented}
Zhaosong Lu and Yong Zhang.
\newblock An augmented lagrangian approach for sparse principal component
  analysis.
\newblock {\em Mathematical Programming}, 135:149--193, 2012.

\bibitem{yuan2013truncated}
Xiao-Tong Yuan and Tong Zhang.
\newblock Truncated power method for sparse eigenvalue problems.
\newblock {\em Journal of Machine Learning Research}, 14(4), 2013.

\bibitem{parkhomenko2007genome}
Elena Parkhomenko, David Tritchler, and Joseph Beyene.
\newblock Genome-wide sparse canonical correlation of gene expression with
  genotypes.
\newblock In {\em BMC proceedings}, volume~1, page S119. Springer, 2007.

\bibitem{hardoon2011sparse}
David~R Hardoon and John Shawe-Taylor.
\newblock Sparse canonical correlation analysis.
\newblock {\em Machine Learning}, 83:331--353, 2011.

\bibitem{mai2019iterative}
Qing Mai and Xin Zhang.
\newblock An iterative penalized least squares approach to sparse canonical
  correlation analysis.
\newblock {\em Biometrics}, 75(3):734--744, 2019.

\bibitem{chen2019alternating}
Shixiang Chen, Shiqian Ma, Lingzhou Xue, and Hui Zou.
\newblock An alternating manifold proximal gradient method for sparse pca and
  sparse cca.
\newblock {\em arXiv preprint arXiv:1903.11576}, 2019.

\bibitem{lindenbaum2021l0}
Ofir Lindenbaum, Moshe Salhov, Amir Averbuch, and Yuval Kluger.
\newblock L0-sparse canonical correlation analysis.
\newblock In {\em International Conference on Learning Representations}, 2021.

\bibitem{lock2013joint}
Eric~F Lock, Katherine~A Hoadley, James~Stephen Marron, and Andrew~B Nobel.
\newblock Joint and individual variation explained (jive) for integrated
  analysis of multiple data types.
\newblock {\em The annals of applied statistics}, 7(1):523, 2013.

\bibitem{tang2021integrated}
Tiffany~M Tang and Genevera~I Allen.
\newblock Integrated principal components analysis.
\newblock {\em Journal of Machine Learning Research}, 22(198):1--71, 2021.

\bibitem{ding2022cooperative}
Daisy~Yi Ding, Shuangning Li, Balasubramanian Narasimhan, and Robert
  Tibshirani.
\newblock Cooperative learning for multiview analysis.
\newblock {\em Proceedings of the National Academy of Sciences},
  119(38):e2202113119, 2022.

\bibitem{ding2023semi}
Daisy~Yi Ding, Xiaotao Shen, Michael Snyder, and Robert Tibshirani.
\newblock Semi-supervised cooperative learning for multiomics data fusion.
\newblock {\em Workshop on Machine Learning for Multimodal Healthcare Data, The
  Fortieth International Conference on Machine Learning (ICML)}, 2023.

\bibitem{strand2022molecular}
Siri~H Strand, Bel{\'e}n Rivero-Guti{\'e}rrez, Kathleen~E Houlahan, Jose~A
  Seoane, Lorraine~M King, Tyler Risom, Lunden~A Simpson, Sujay Vennam, Aziz
  Khan, Luis Cisneros, et~al.
\newblock Molecular classification and biomarkers of clinical outcome in breast
  ductal carcinoma in situ: Analysis of tbcrc 038 and rahbt cohorts.
\newblock {\em Cancer Cell}, 40(12):1521--1536, 2022.

\bibitem{hand2001simple}
David~J Hand and Robert~J Till.
\newblock A simple generalisation of the area under the roc curve for multiple
  class classification problems.
\newblock {\em Machine Learning}, 45:171--186, 2001.

\bibitem{schmidt2021prognostic}
Marcus Schmidt, Karolina Edlund, Jan~G Hengstler, Anne-Sophie Heimes, Katrin
  Almstedt, Antje Lebrecht, Slavomir Krajnak, Marco~J Battista, Walburgis
  Brenner, Annette Hasenburg, et~al.
\newblock Prognostic impact of immunoglobulin kappa c (igkc) in early breast
  cancer.
\newblock {\em Cancers}, 13(14):3626, 2021.

\bibitem{schmidt2012immunoglobulin}
Marcus Schmidt, Patrick Micke, Mathias Gehrmann, and Jan~G Hengstler.
\newblock Immunoglobulin kappa chain as an immunologic biomarker of prognosis
  and chemotherapy response in solid tumors.
\newblock {\em Oncoimmunology}, 1(7):1156--1158, 2012.

\bibitem{prest2002estrogen}
Sara~J Prest, Felicity~EB May, and Bruce~R Westley.
\newblock The estrogen-regulated protein, tff1, stimulates migration of human
  breast cancer cells.
\newblock {\em The FASEB Journal}, 16(6):592--594, 2002.

\bibitem{perry2009cross}
Patrick~O Perry.
\newblock {\em Cross-validation for unsupervised learning}.
\newblock Stanford University, 2009.

\end{thebibliography}
\bibliographystyle{unsrt} %unsrt

\newpage
\begin{appendix}
\section{Cross-validation procedure for CoCA}
\label{sec:appendix_cv}

In this section, we describe three options for the cross-validation (CV) procedure to determine the optimal values of hyperparameters $\rho$ and $\lambda$ in CoCA: $K$-fold CV and ``speckled CV'' for unsupervised settings, and $K$-fold CV for supervised settings.

For $K$-fold CV in unsupervised settings, the training set is divided into $K$ folds. 
For each combination of $\rho$ and $\lambda$ in a pre-defined grid, we iterate through the folds. In each iteration, we train CoCA on $K-1$ folds, excluding the current fold, and then calculate the reconstruction error on the held-out fold using the estimated component. 
This process is repeated $K$ times, and the average reconstruction error across all $K$ folds is computed for each hyperparameter combination. 
The $\rho$ and $\lambda$ values that minimize this average reconstruction error across folds are selected as optimal. 

An alternative approach is ``speckled CV'', where a proportion of values in the data matrix are randomly masked. 
These masked values serve as a validation set for hyperparameter selection, while the unmasked values are used for training.
In this approach, we evaluate how well the component derived from the unmasked data can reconstruct the masked elements. 
Specifically, for each combination of hyperparameters, we compute the components using the unmasked data, reconstruct the entire matrix, and then calculate the reconstruction error for the masked elements. 
The hyperparameters that minimize this reconstruction error are selected as optimal. 
This procedure is described in more detail in \cite{perry2009cross}.

When an outcome of interest is available and is desired to be used for hyperparameter selection we follow a similar $K$-fold CV procedure, but with a focus on predicting the outcome of interest rather than reconstruction error. 
For each hyperparameter combination, we apply CoCA on the training data and derive multi-view scores.
We then use $K$-1 folds of these scores to predict the responses in the $K$-th fold. 
This process is repeated $K$ times, with each fold serving as the validation set once.
We calculate a prediction error metric (such as misclassification rate or mean squared error) for each fold. The average prediction error across all folds is computed for each hyperparameter combination, and the $\rho$ and $\lambda$ values that minimize this average prediction error are selected as optimal.

\section{Proof of Theorem~\ref{thm:coca-solution-path}}
\label{sec:appendix_relation}

We begin by establishing that the solutions $\hat{u},\hat{v}$ to Problem~\eqref{eqn:cooperative-svd-2} satisfy~\eqref{eqn:cooperative-svd-u} and~\eqref{eqn:cooperative-svd-v}, using standard SVD arguments. Define
$$
\mc{L}(u,v) := \frac{1}{2}v^{\top}v - u^{\top} \bX v + \frac{\rho}{2}\|\bX_1 v_1 - \bX_2 v_2\|_2^2.
$$
This is simply the criterion of Problem~\eqref{eqn:cooperative-svd-2} minus $\tr(\bX^{\top} \bX)$, and it follows that Problem~\eqref{eqn:cooperative-svd-2} is equivalent to
\begin{equation}
    \label{eqn:cooperative-svd-3}
    \min_{u,v} \mc{L}(u,v), \quad \st \|u\|_2^2 = 1.
\end{equation}
Let $\hat{v}(u)$ be the solution to $\underset{v}{\min}\mc{L}(u,v)$, and observe that
\begin{equation}
\label{eqn:cooperative-svd-4}
\bX^{\top} u = (\Identity + \rho \bD \bX^{\top} \bX \bD) \hat{v}(u).
\end{equation}
Substituting this expression for $v$ in~\eqref{eqn:cooperative-svd-3}, it follows that $\hat{u}$ is the solution to
\begin{equation*}
\min_u -u^{\top} \bX (\Identity + \rho \bD \bX^{\top} \bX \bD)^{-1} \bX^{\top} u, \quad \st \|u\|_2^2 = 1.
\end{equation*}
Therefore $\hat{u}$ satisfies~\eqref{eqn:cooperative-svd-u} as claimed:
\begin{equation}
\label{eqn:cooperative-svd-u-appendix}
\bX (\Identity + \rho \bD \bX^{\top} \bX \bD)^{-1} \bX^{\top} \hat{u} = \lambda_1 \hat{u}, \quad \|\hat{u}\|_2 = 1.
\end{equation} 
Now, let $\hat{u}(v)$ solve $\min_{u} \mc{L}(u,v) \st \|u\|_2^2 = 1$. Observe that 
\begin{equation}
\label{eqn:cooperative-svd-uv}
\hat{u}(v) = \frac{\bX v}{\|\bX v\|_2}.
\end{equation}
Plugging this in to~\eqref{eqn:cooperative-svd-u} and applying $\bX^{\top}$ to each side gives:
\begin{align*}
\frac{1}{\|\bX \hat{v}\|_2}\bX^{\top} \bX (\Identity + \rho \bD \bX^{\top} \bX \bD)^{-1} \bX^{\top} \bX \hat{v} & = \frac{1}{\|\bX \hat{v}\|_2}\lambda_1 \bX^{\top} \bX \hat{v}, \Longleftrightarrow \\
(\Identity + \rho \bD \bX^{\top} \bX \bD)^{-1} \bX^{\top} \bX \hat{v} & = \lambda_1 \hat{v}.
\end{align*}
We have now solved for $\hat{u}$ and $\hat{v}$, the latter up to a constant of proportionality. To determine what this constant is, note that from~\eqref{eqn:cooperative-svd-4},~\eqref{eqn:cooperative-svd-uv} and the previous display we have that
\begin{equation*}
\hat{v} = (\Identity + \rho \bD \bX^{\top} \bX \bD)^{-1} \bX^{\top} \hat{u} = \frac{1}{\|\bX \hat{v}\|_2}(\Identity + \rho \bD \bX^{\top} \bX \bD)^{-1} \bX^{\top} \bX \hat{v} = \frac{\lambda_1}{\|\bX \hat{v}\|_2} \hat{v}.
\end{equation*}
In other words, $\hat{v}$ is proportional to the leading eigenvector of the matrix $(\Identity + \rho \bD \bX^{\top} \bX \bD)^{-1} \bX^{\top} \bX$, and has variance equal to the square of the leading eigenvalue of this matrix:
\begin{equation}
\label{eqn:cooperative-svd-v-appendix}
(\Identity + \rho \bD \bX^{\top} \bX \bD)^{-1} \bX^{\top} \bX \hat{v} = \lambda_1 \hat{v}, \quad \|\bX \hat{v}\|_2 = \lambda_1.
\end{equation}
Now we establish the statements in Theorem~\ref{thm:coca-solution-path} regarding the equivalence of CoCA to PCA and CCA at the two ends of its solution path. The equivalence to PCA is true by definition -- that is, when \textbf{$\rho = 0$} by definition $\hat{v}$ is proportional to the first principal component of $\bX$ -- and so we concentrate on deriving the equivalence to CCA as $\rho \to \infty$. 

Since we have assumed $\rank(\bX) = p$, $\bX^{\top} \bX$ is invertible. In that case, note that, as $\rho \to \infty$, $\hat{v}/\|\hat{v}\|_2$ converges to the leading eigenvector of the matrix $(\bD \bX^{\top} \bX \bD)^{-1} \bX^{\top} \bX$. The eigenpairs of this matrix are the solutions $(v_k,\lambda_k), k = 1,\ldots,p$ to the eigenproblem
\begin{equation}
    \label{eqn:eigenproblem-cca}
    (\bD \bX^{\top} \bX \bD)^{-1} \bX^{\top} \bX v = \lambda v, \quad \|v\|^2 = 1.
\end{equation}
We begin by showing that each pair of canonical directions between $\bX_1$ and $\bX_2$ defines an eigenvector of $(\bD \bX^{\top} \bX \bD)^{-1} \bX^{\top} \bX v$. Recall that there are a total of $p^* := \min\{p_1,p_2\}$ canonical pairs $(\tilde{v}_{1},\tilde{v}_{2},\gamma)$ between $\bX_1$ and $\bX_2$, each of which satisfy the stationary conditions
\begin{equation}
\begin{aligned}
    \label{eqn:cca}
    (\bX_1^{\top} \bX_2) \tilde{v}_2 & = \sqrt{\gamma} (\bX_1^{\top} \bX_1) \tilde{v}_1,  \quad \tilde{v}_1^{\top} \bX_1^{\top} \bX_1 \tilde{v}_1 = 1 \\
    (\bX_2^{\top} \bX_1) \tilde{v}_1 & = \sqrt{\gamma} (\bX_2^{\top} \bX_2) \tilde{v}_2, \quad \tilde{v}_2^{\top} \bX_2^{\top} \bX_2 \tilde{v}_2 = 1,
\end{aligned}
\end{equation}
where $\gamma$ defines the squared cross-correlation between $\tilde{v}_1$ and $\tilde{v}_2$. We will assume for ease of exposition that the cross-correlations have multiplicity $1$, so that~\eqref{eqn:cca} defines $(\tilde{v_1},\tilde{v}_2,\gamma)$ up to the sign of $\tilde{v}_1,\tilde{v}_2$. To see the relationship between the eigenvectors $v_k$ and canonical directions $\tilde{v}_1,\tilde{v}_2$, notice that the eigenvector equation~\eqref{eqn:eigenproblem-cca} can be rearranged to 
$$
\bD \bX^{\top} \bX \bD v  = \frac{1}{\lambda} \bX^{\top} \bX v.
$$
Written in block-matrix form, this is
\begin{align*}
\bX_1^{\top} \bX_1 v_1  - \bX_1^{\top} \bX_2 v_2
& = \frac{1}{\lambda} (\bX_1^{\top} \bX_1 v_1 + \bX_1^{\top} \bX_2 v_2) \\
-\bX_2^{\top} \bX_1 v_1  + \bX_2^{\top} \bX_2 v_2
& = \frac{1}{\lambda} (\bX_2^{\top} \bX_1 v_1 + \bX_2^{\top} \bX_2 v_2),
\end{align*}
which can be rearranged to read
\begin{equation}
\begin{aligned}
\label{eqn:cca-2}
    (\bX_1^{\top} \bX_2) {v}_2 & = \frac{(1 - \lambda^{-1})}{(1 + \lambda^{-1})} (\bX_1^{\top} \bX_1) {v}_1 \\
    (\bX_2^{\top} \bX_1) {v}_1 & = \frac{(1 - \lambda^{-1})}{(1 + \lambda^{-1})} (\bX_2^{\top} \bX_2) {v}_2,
\end{aligned}
\end{equation}
This is simply~\eqref{eqn:cca} with $\sqrt{\gamma} = \frac{(1 - \lambda^{-1})}{(1 + \lambda^{-1})}$. It follows that each canonical pair corresponds to an eigenpair of~\eqref{eqn:eigenproblem-cca}, meaning precisely that if $(\tilde{v}_1,\tilde{v}_2,\gamma)$ is a canonical pair, then (letting $\tilde{v} = (\tilde{v}_1,\tilde{v}_2)$)
$$
\bigg(\frac{\tilde{v}}{\|\tilde{v}\|_2}, \frac{1 + \sqrt{\gamma}}{1 - \sqrt{\gamma}}\bigg)
$$
is an eigenpair of~\eqref{eqn:eigenproblem-cca}. Additionally, notice that if $(v,\lambda)$ is an eigenpair satisfying~\eqref{eqn:eigenproblem-cca}, then $(\bD v,1/\lambda)$ is also an eigenpair satisfying~\eqref{eqn:eigenproblem-cca}. As a result, 
$$
\bigg(\frac{\bD\tilde{v}}{\|\tilde{v}\|_2}, \frac{1 - \sqrt{\gamma}}{1 + \sqrt{\gamma}}\bigg)
$$
is also an eigenpair of~\eqref{eqn:eigenproblem-cca}. 

This characterizes $2p^{\ast}$ eigenpairs of~\eqref{eqn:eigenproblem-cca}. The remaining $p - 2p^{\ast}$ eigenpairs correspond to the $p_1 - p_2$-dimensional subspace of $\col(\bX_1)$ that lies in the null space of $\bX_2^{\top}$ (assuming without loss of generality that $p_2 < p_1$); for any such eigenpair $(v,\lambda)$ it is the case that $v_2 = 0$ and $\lambda = 1$. None of these are the leading eigenpair, since $\bX_1^{\top} \bX_2 \neq 0$ and therefore $\lambda_1 > 1$. We conclude that the leading canonical pair $(\tilde{v}_{11},\tilde{v}_{12},\gamma_1)$ corresponds to leading eigenpair of $(\bD \bX^{\top} \bX \bD)^{-1} \bX^{\top} \bX$. This is the desired claim.

\section{Alternative formulations of CoCA}
\label{sec:other_formulation}
%{\color{blue} To do for Daisy: update this section}

As mentioned, we have experimented with alternative formulations of CoCA.

\paragraph{Alternative formulation 1} One alternative formulation is to minimize the following objective:
\begin{equation}
 {\min_{u,v}} \; \frac12 ||\bX-\theta\cdot \bX vu^{\top} ||_F^2 +\frac{\rho}{2} ||\bX_1v_1-\bX_2v_2||_2^2 + \lambda ||v||_1, \; \textrm{subject to}\; ||u||_2^2 = 1, ||v||_2^2 = 1.
 \label{eq_cpc_sparse_1}
\end{equation}

Here $\rho \geq 0$ controls the relative importance of the agreement penalty and $\lambda \geq 0$ controls the level of sparsity.
$v \in \Reals^{p}$ and $u \in \Reals^{n}$ are vectors, and $\theta$ is a scalar.
$v$ is partitioned as $v = (v_1, v_2)$, where $v_1 \in \Reals^{p_1}$ corresponds to $\bX_1$ and $v_2 \in \Reals^{p_2}$ to $\bX_2$. However, this formulation has an important drawback: as $\rho$ approaches infinity, the solution does not converge to the CCA solution. 

\paragraph{Alternative formulation 2}
The limitation of \eqref{eq_cpc_sparse_1} prompted us to explore another alternative formulation:
\begin{equation}
 {\min_{\theta, u,v}} \; \frac12 ||\bX-\theta\cdot \bX vu^{\top} ||_F^2 +\frac{\rho}{2} ||\bX_1 v_1-\bX_2 v_2||_2^2 + \lambda ||v||_1, \; \textrm{subject to}\; ||u||^2_2\leq 1, ||\bX v||^2_2 = 1.
 \label{eq_cpc_sparse}
\end{equation}

Setting $\lambda = 0$, it can be shown that this formulation corresponds to PCA when $\rho = 0$ and CCA as $\rho \to \infty$; we omit the derivation as is it similar to the proof of Theorem~\ref{thm:coca-solution-path}.

The difficulty with this formulation is that constraint $||\bX v||^2_2 = 1$ makes the problem challenging to solve. This can be finessed in the following way. We begin from~\eqref{eq_cpc_sparse}, without sparsity:
\begin{equation}
 {\min_{\theta, u,v}} \; \frac12 ||\bX-\theta\cdot \bX vu^{\top} ||_F^2 +\frac{\rho}{2} ||\bX_1 v_1-\bX_2 v_2||_2^2, \; \textrm{subject to}\; ||u||^2_2\leq 1, ||\bX v||^2_2 = 1.
 \label{eq_cpc}
\end{equation}
First, solving explicitly for $\theta, u$, we see that \eqref{eq_cpc} is equivalent to the following optimization problem:
\begin{equation*}
	\max_{v} v^{\top} \bSigma^2 v - \rho v^{\top} \DifferenceMatrix \bSigma \DifferenceMatrix v, \quad \textrm{subject to}~~v^{\top} \bSigma v = 1.
\end{equation*}
By adding $c \bSigma$ for a sufficiently large value of $c$--for instance taking $c$ to be the maximum eigenvalue of $\rho \DifferenceMatrix \bSigma \DifferenceMatrix$ will suffice--we can reformulate this as the generalized eigenproblem
\begin{equation}
	\label{eqn:coop-pca-with-psd-objective}
	\max_{v} v^{\top} {\boldsymbol M}_{\rho} v, \quad \textrm{subject to}~~v^{\top} \bSigma v = 1,
\end{equation}
 where ${\boldsymbol M}_{\rho} = \bSigma^2 - \rho v^{\top} \DifferenceMatrix \bSigma \DifferenceMatrix + c\Identity$ is PSD. Problem~\eqref{eqn:coop-pca-with-psd-objective} is equivalent to the following convex relaxation,
 \begin{equation*}
 	\min_{\bV \in \Reals^{p \times p}} -\tr(\bM_{\rho}\bV), \quad \textrm{subject to}~~ \tr(\bSigma\bV) = 1, \bV \succeq 0,
 \end{equation*}
in the sense that the solution will be the rank-1 matrix $\hat{\bV} = v_1 v_1^{\top}$ where $v_1$ is the leading eigenvector of $\bM_{\rho}$. 

Now, to incorporate sparsity, we constrain the $\ell^1$ norm of $\bV$ to be at most $k^2$:
\begin{equation}
	\label{eqn:sparse-coop-pca}
	\min_{\bV \in \Reals^{p \times p}} -\tr({\boldsymbol M}_{\rho}\bV), \quad \textrm{subject to}~~\tr(\bSigma \bV) = 1, \bV \succeq 0, \|\bV\|_1 \leq k^2. 
\end{equation}
The problem~\eqref{eqn:sparse-coop-pca} is a semidefinite program (SDP), and can either be solved exactly using standard SDP solvers, or approximately using a first-order method. However, it should be noted that the solution is no longer guaranteed to be rank-$1$. Additionally, solving semidefinite programs becomes practically difficult for large-scale problems. Instead, we opt for the CoCA formulation~\eqref{eqn:sparse-cooperative-svd} presented earlier.
This formulation satisfies our key requirements: it encompasses PCA and CCA as special cases at the extremes of the regularization path, and it remains computationally tractable when incorporating sparsity, allowing for efficient optimization using algorithms like \texttt{glmnet} in an alternating scheme.

\end{appendix}

\end{document}